\documentclass[preprint2]{aastex}

\slugcomment{The Astrophysical Journal}

\shorttitle{TeV and MWL Observations of Mrk\,421}
\shortauthors{The VERITAS Collaboration and MWL partners}

\begin{document}



\title{TeV and Multi-wavelength Observations of Mrk\,421 in 2006-2008}



\author{
V.~A.~Acciari\altaffilmark{1},
E.~Aliu\altaffilmark{2},
T.~Arlen\altaffilmark{3},
T.~Aune\altaffilmark{4},
M.~Beilicke\altaffilmark{5,\ast},
W.~Benbow\altaffilmark{1},
D.~Boltuch\altaffilmark{6},
S.~M.~Bradbury\altaffilmark{7},
J.~H.~Buckley\altaffilmark{5},
V.~Bugaev\altaffilmark{5},
K.~Byrum\altaffilmark{8},
A.~Cannon\altaffilmark{9},
A.~Cesarini\altaffilmark{10},
L.~Ciupik\altaffilmark{11},
W.~Cui\altaffilmark{12},
R.~Dickherber\altaffilmark{5},
C.~Duke\altaffilmark{13},
A.~Falcone\altaffilmark{14},
J.~P.~Finley\altaffilmark{12},
G.~Finnegan\altaffilmark{15},
L.~Fortson\altaffilmark{11},
A.~Furniss\altaffilmark{4},
N.~Galante\altaffilmark{1},
D.~Gall\altaffilmark{19},
G.~H.~Gillanders\altaffilmark{10},
S.~Godambe\altaffilmark{15},
J.~Grube\altaffilmark{11},
R.~Guenette\altaffilmark{16},
G.~Gyuk\altaffilmark{11},
D.~Hanna\altaffilmark{16},
J.~Holder\altaffilmark{6},
C.~M.~Hui\altaffilmark{15},
T.~B.~Humensky\altaffilmark{17},
A.~Imran\altaffilmark{18},
P.~Kaaret\altaffilmark{19},
N.~Karlsson\altaffilmark{11},
M.~Kertzman\altaffilmark{20},
D.~Kieda\altaffilmark{15},
A.~Konopelko\altaffilmark{21},
H.~Krawczynski\altaffilmark{5},
F.~Krennrich\altaffilmark{18},
M.~J.~Lang\altaffilmark{10},
G.~Maier\altaffilmark{16,\amalg},
S.~McArthur\altaffilmark{5},
M.~McCutcheon\altaffilmark{16},
P.~Moriarty\altaffilmark{22},
R.~A.~Ong\altaffilmark{3},
A.~N.~Otte\altaffilmark{4},
M.~Ouellette\altaffilmark{23},
D.~Pandel\altaffilmark{19},
J.~S.~Perkins\altaffilmark{1},
A.~Pichel\altaffilmark{24},
M.~Pohl\altaffilmark{18,\mho},
J.~Quinn\altaffilmark{9},
K.~Ragan\altaffilmark{16},
L.~C.~Reyes\altaffilmark{25},
P.~T.~Reynolds\altaffilmark{26},
E.~Roache\altaffilmark{1},
H.~J.~Rose\altaffilmark{7},
A.~C.~Rovero\altaffilmark{24},
M.~Schroedter\altaffilmark{18},
G.~H.~Sembroski\altaffilmark{12},
G.~Demet~Senturk\altaffilmark{27},
D.~Steele\altaffilmark{11,\diamond},
S.~P.~Swordy\altaffilmark{17},
M.~Theiling\altaffilmark{1},
S.~Thibadeau\altaffilmark{5},
A.~Varlotta\altaffilmark{12},
V.~V.~Vassiliev\altaffilmark{3},
S.~Vincent\altaffilmark{15},
R.~G.~Wagner\altaffilmark{8},
S.~P.~Wakely\altaffilmark{17},
J.~E.~Ward\altaffilmark{9},
T.~C.~Weekes\altaffilmark{1},
A.~Weinstein\altaffilmark{3},
T.~Weisgarber\altaffilmark{17},
D.~A.~Williams\altaffilmark{4},
S.~Wissel\altaffilmark{17},
M.~Wood\altaffilmark{3},
B.~Zitzer\altaffilmark{12}
}

\altaffiltext{1}{Fred Lawrence Whipple Observatory, Harvard-Smithsonian 
Center for Astrophysics, Amado, AZ 85645, USA}
\altaffiltext{2}{Department of Physics and Astronomy, Barnard College, 
Columbia University, NY 10027, USA}
\altaffiltext{3}{Department of Physics and Astronomy, University of 
California, Los Angeles, CA 90095, USA}
\altaffiltext{4}{Santa Cruz Institute for Particle Physics and 
Department of Physics, University of California, Santa Cruz, CA 95064, 
USA}
\altaffiltext{5}{Department of Physics, Washington University, St. 
Louis, MO 63130, USA}
\altaffiltext{6}{Department of Physics and Astronomy and the Bartol 
Research Institute, University of Delaware, Newark, DE 19716, USA}
\altaffiltext{7}{School of Physics and Astronomy, University of Leeds, 
Leeds, LS2 9JT, UK}
\altaffiltext{8}{Argonne National Laboratory, 9700 S. Cass Avenue, 
Argonne, IL 60439, USA}
\altaffiltext{9}{School of Physics, University College Dublin, Belfield, 
Dublin 4, Ireland}
\altaffiltext{10}{School of Physics, National University of Ireland 
Galway, University Road, Galway, Ireland}
\altaffiltext{11}{Astronomy Department, Adler Planetarium and Astronomy 
Museum, Chicago, IL 60605, USA}
\altaffiltext{12}{Department of Physics, Purdue University, West 
Lafayette, IN 47907, USA }
\altaffiltext{13}{Department of Physics, Grinnell College, Grinnell, IA 
50112-1690, USA}
\altaffiltext{14}{Department of Astronomy and Astrophysics, 525 Davey 
Lab, Pennsylvania State University, University Park, PA 16802, USA}
\altaffiltext{15}{Department of Physics and Astronomy, University of 
Utah, Salt Lake City, UT 84112, USA}
\altaffiltext{16}{Physics Department, McGill University, Montreal, QC 
H3A 2T8, Canada}
\altaffiltext{17}{Enrico Fermi Institute, University of Chicago, 
Chicago, IL 60637, USA}
\altaffiltext{18}{Department of Physics and Astronomy, Iowa State 
University, Ames, IA 50011, USA}
\altaffiltext{19}{Department of Physics and Astronomy, University of 
Iowa, Van Allen Hall, Iowa City, IA 52242, USA}
\altaffiltext{20}{Department of Physics and Astronomy, DePauw 
University, Greencastle, IN 46135-0037, USA}
\altaffiltext{21}{Department of Physics, Pittsburg State University, 
1701 South Broadway, Pittsburg, KS 66762, USA}
\altaffiltext{22}{Department of Life and Physical Sciences, Galway-Mayo 
Institute of Technology, Dublin Road, Galway, Ireland}
\altaffiltext{23}{Physics Department, California Polytechnic State 
University, San Luis Obispo, CA 94307, USA}
\altaffiltext{24}{Instituto de Astronomia y Fisica del Espacio, Casilla 
de Correo 67 - Sucursal 28, (C1428ZAA) Ciudad Autónoma de Buenos Aires, 
Argentina}
\altaffiltext{25}{Kavli Institute for Cosmological Physics, University 
of Chicago, Chicago, IL 60637, USA}
\altaffiltext{26}{Department of Applied Physics and Instrumentation, 
Cork Institute of Technology, Bishopstown, Cork, Ireland}
\altaffiltext{27}{Columbia Astrophysics Laboratory, Columbia University, 
New York, NY 10027, USA}

\altaffiltext{$\amalg$}{Now at DESY, Platanenallee 6, 15738 Zeuthen, 
Germany}
\altaffiltext{$\mho$}{Now at Institut f\"{u}r Physik und Astronomie, 
Universit\"{a}t Potsdam, 14476 Potsdam-Golm,Germany; DESY, Platanenallee 
6, 15738 Zeuthen, Germany}
\altaffiltext{$\diamond$}{Now at Los Alamos National Laboratory, MS 
H803, Los Alamos, NM 87545}


\author{
A.~Garson~III\altaffilmark{5},
K.~Lee\altaffilmark{5} ---
BRT/NMS: A.C.~Sadun\altaffilmark{101} ---
Bell: M.~Carini\altaffilmark{102},
D.~Barnaby\altaffilmark{102},
K.~Cook\altaffilmark{102},
J.~Maune\altaffilmark{102},
A.~Pease\altaffilmark{102},
S.~Smith\altaffilmark{102},
R.~Walters\altaffilmark{102} ---
%
Tuorla/KVA: A.~Berdyugin\altaffilmark{103},
E.~Lindfors\altaffilmark{103},
K.~Nilsson\altaffilmark{103},
M.~Pasanen\altaffilmark{103},
J.~Sainio\altaffilmark{103},
A.~Sillanpaa\altaffilmark{103},
L.O.~Takalo\altaffilmark{103},
C.~Villforth\altaffilmark{103} ---
%
WIYN: T.~Montaruli\altaffilmark{104},
M.~Baker\altaffilmark{104} ---
%
Mets\"ahovi: A.~Lahteenmaki\altaffilmark{105},
M.~Tornikoski\altaffilmark{105},
T.~Hovatta\altaffilmark{105},
E.~Nieppola\altaffilmark{105} ---
%
UMRAO: H.D.~Aller\altaffilmark{106},
M.F.~Aller\altaffilmark{106}
}

\altaffiltext{101}{University of Colorado, Denver CO 80217-3364, USA}

\altaffiltext{102}{Western Kentucky University, Physics and Astronomy,
1906 College Heights Blvd, Bowling Green, KY 42103, USA}

\altaffiltext{103}{Tuorla Observatory, Department of Physics and
Astronomy, University of Turku, Finland}

\altaffiltext{104}{University of Wisconsin - Madison, Physics
Department, 53706, Madison, WI}

\altaffiltext{105}{Aalto University Metsahovi Radio Observatory,
Metsahovintie 114, FIN-02540 Kylmala, Finland}

\altaffiltext{106}{Department of Astronomy, University of Michigan, Ann
Arbor, MI 48109-1042, USA}

\altaffiltext{$\ast$}{Corresponding author: beilicke@physics.wustl.edu}



\begin{abstract}

We report on TeV $\gamma$-ray observations of the blazar Mrk\,421
(redshift of $0.031$) with the VERITAS observatory and the Whipple $10
\, \rm{m}$ Cherenkov telescope. The excellent sensitivity of VERITAS
allowed us to sample the TeV $\gamma$-ray fluxes and energy spectra with
unprecedented accuracy where Mrk\,421 was detected in each of the
pointings. A total of $47.3 \, \rm{hrs}$ of VERITAS and $96 \, \rm{hrs}$
of Whipple $10 \, \rm{m}$ data were acquired between January 2006 and
June 2008. We present the results of a study of the TeV $\gamma$-ray
energy spectra as a function of time, and for different flux levels. On
May 2nd and 3rd, 2008, bright TeV $\gamma$-ray flares were detected with
fluxes reaching the level of 10~Crab. The TeV $\gamma$-ray data were
complemented with radio, optical, and X-ray observations, with flux
variability found in all bands except for the radio waveband. The
combination of the RXTE and Swift X-ray data reveal spectral hardening
with increasing flux levels, often correlated with an increase of the
source activity in TeV $\gamma$-rays. Contemporaneous spectral energy
distributions were generated for 18 nights, each of which are reasonably
described by a one-zone SSC model.

\end{abstract}


\keywords{BL Lacertae objects: individual (Mrk\,421) --- Galaxies: jets
--- Galaxies: nuclei --- Gamma rays: galaxies --- X-rays: galaxies}

\section{Introduction \label{sec:introduction}}

In 1992, observations with the Whipple $10 \, \rm{m}$ Cherenkov
telescope led to the first discovery of an extragalactic source of TeV
$\gamma$-rays, the blazar Mrk\,421 \citep{Punch1992}. Since then, more
than 30 similar sources have been detected with ground-based
$\gamma$-ray detectors \citep{TeVCat}. The sources with well measured
red shifts lie from $0.031$ (Mrk\,421, \citet{Mrk421Redshift}) to
$0.536$ for the recently detected radio quasar 3C\,279 \citep{alb08}.
Typically, blazars show core-dominated emission, and they are
characterized by rapid variability. Their spectral energy distribution
(SED) in the $\nu F_{\nu}$ representation is characterized by two broad,
well-separated ``humps'' arising from (i) synchrotron emission
(low-energy) and (ii) a high-energy component of either leptonic or
hadronic nature. During TeV $\gamma$-ray flares, strong sources (i.e.
Mrk\,421 and PKS\,2155-304) exhibit $\nu F_{\nu}$-fluxes of about
$10^{-9} \, \rm{erg} \, \rm{cm}^{-2} \, \rm{s}^{-1}$ \citep{PKS2155};
corresponding to $\gamma$-ray luminosities of between $10^{42}$ and
$10^{43} \, \rm{erg} \, \rm{s}^{-1}$ for assumed anisotropic emission
with an opening angle of $5 \deg$. The blazars detected at TeV energies
are the high-frequency peaked counterparts of the blazar source
population detected at MeV/GeV energies with the {\it EGRET} experiment
on board the space-borne {\it Compton Gamma Ray Observatory}
\citep{Hartman1999}, and recently expanded by the {\it Large Area
Telescope} (LAT) on board the Fermi $\gamma$-ray satellite
\citep{FirstFermiCatalog}.

Following the detection in TeV $\gamma$-rays in 1992, Mrk\,421 was
observed intensively, and the observations led to a number of landmark
discoveries:

\begin{enumerate}

\item The detection of fast flux variability with a doubling time of $15
\, \rm{min}$ \citep{Gaidos1996}.

\item The first tentative evidence for a X-ray/TeV $\gamma$-ray flux
correlation \citep{Buckley1996}. Several observation campaigns
strengthened the evidence for such a correlation. Some of the most
convincing results were presented by \citet{Fossati2008}.

\item The detection of TeV $\gamma$-ray energy spectra that harden with
increasing fluxes \citep{Krennrich2002}.

\end{enumerate}   

Modeling of Mrk\,421 data with synchrotron-Compton models revealed the
first evidence for bulk jet Lorentz factors of the order of 50
\citep{Krawczynski2001} and modeling of data taken during different
states revealed evidence that one-zone synchrotron self-Compton (SSC)
models are insufficient to describe the observations
\citep{Blazejovski2005}. Alternative models are discussed in the
literature. Recent papers include \citet{Katarzynski2010, Boettcher2010,
Gao2010, Tammi2009, Lichti2008, Stecker2007}. These models involve more
complicated geometries which lead to a larger number of free model
parameters than in the SSC model.

Mrk\,421 has been a frequent target of multi-wavelength campaigns. During
some of its very short flares the X-ray and TeV $\gamma$-ray fluxes
tracked each other \citep{Fossati2008}. However, X-ray flares that are
not accompanied by TeV $\gamma$-ray flares and vice versa have also been
observed \citep{Rebillot2006,Fossati2008}. There is good evidence that
the X-ray and TeV $\gamma$-ray activities are correlated when averaged
over $\sim$1 week time intervals \citep{Blazejovski2005,Horan2009}. All
attempts to establish convincing evidence for a correlation of the X-ray
and $\gamma$-ray fluxes with the flux variability at radio to optical
wavelengths, have failed so far \citep{Blazejovski2005,Horan2009}.

The X-ray/TeV $\gamma$-ray correlation properties are of great interest,
as they might enable us to decide on the emission model, i.e. between
(i) leptonic models in which a single population of high energy
electrons emits the low and high-energy radiation as synchrotron and
inverse Compton (IC) emission, respectively, and (ii) hadronic models in
which the $\gamma$-ray emission is attributed to additional
population(s) of high-energy particles, powered by the acceleration of
extremely high-energy protons. It is interesting to note that the
X-ray/TeV $\gamma$-ray flux correlation has only been studied for a
handful of sources (Mrk\,421, Mrk\,501, 1\,ES1959+650, PKS\,2155-305,
1\,ES2344+514) with a sufficiently good signal-to-noise ratio in both
bands to investigate the correlation properties. Although the X-ray and
TeV $\gamma$-ray fluxes seem to be correlated, it is not clear how well
this correlation holds for individual flares
\citep{Krawczynski_1ES1959}.

In this paper we report on the first TeV $\gamma$-ray observation
campaign on Mrk\,421 performed with the VERITAS observatory. VERITAS
achieves an energy flux sensitivity of $\left(t_{\rm obs}/50\rm
hrs\right)^{-1/2} \times 5 \times 10^{-13} \, \rm{ergs} \, \rm{cm}^{-2}
\, \rm{sec}^{-1}$ at $500 \, \rm{GeV}$ \footnote{In terms of the flux
from the Crab Nebula, the VERITAS sensitivity is $\left( t_{\rm{obs}} /
30 \, \rm{min} \right)^{-1/2}$ $\times$ 8\% Crab in $30 \, \rm{min}$ of
observations. This number is valid for the VERITAS array as operated in
2008.}. The high sensitivity of VERITAS and the brightness of Mrk\,421
during flares allow us to measure fluxes with $\sim$minute time bins,
and to determine energy spectra for 5-min time intervals or less.  We
also report accompanying observations in the radio band, at optical
wavelengths, and in the X-ray band with the {\it Rossi X-ray Timing
Explorer} (RXTE) PCA, {\it Swift} XRT and {\it Suzaku} instruments.

In Section~\ref{sec:datasets} we discuss the data sets and the analysis 
methods. The results of different studies are presented in 
Section~\ref{sec:results} followed by a summary and discussion in 
Section~\ref{sec:discussion}.

\section{Data Sets and Data Reduction \label{sec:datasets}}

In this section we describe the observations and analysis of the data 
taken in the TeV energy regime (Section~\ref{subsec:DataVHE}), at X-ray 
energies (Section~\ref{subsec:DataXRay}), as well as in the optical 
(Section~\ref{subsec:DataOptical}) and radio 
(Section~\ref{subsec:DataRadio}) wavebands.

\subsection{VERITAS/Whipple $\gamma$-ray data}
\label{subsec:DataVHE}

\paragraph{VERITAS.}

VERITAS\footnote{Very Energetic Radiation Imaging Telescope Array
System.} consists of four $12 \, \rm{m}$ diameter imaging atmospheric
Cherenkov telescopes (IACTs) and is located at the base camp of the Fred
Lawrence Whipple Observatory (FLWO) in southern Arizona at an altitude
of $1280 \, \rm{m}$. It detects the Cherenkov light emitted by an
extensive air shower (initiated by a $\gamma$-ray photon or cosmic ray
entering the Earth's atmosphere) using a 499-pixel photomultiplier
camera located in the focal plane of each telescope. The array is
sensitive to $\gamma$-rays in the energy range from $\sim$100~GeV to
$\sim$30~TeV. Observations are performed (in $20 \, \rm{min}$ runs) on
moonless nights using the ``wobble'' mode of operation, where all
telescopes are pointed to a sky position offset of $\pm 0.5 \deg$
(alternating in direction between consecutive data runs) with respect to
the source position. This method allows for a simultaneous background
estimation to be made. More details about VERITAS, the data calibration
and the analysis techniques can be found in \citet{acc08}.

Only shower images which pass certain quality cuts are considered in the
event reconstruction: image size $\geq 500$ digital counts\footnote{The
photomultiplier pulses are integrated within a time window of $24 \,
\rm{ns}$ duration. One photoelectron corresponds to approximately $5$
digital counts.} (dc) and image distance to the center of the camera
$<$1.43~$\deg$. The standard cuts for $\gamma$/hadron separation, which
are based on the width and length of the recorded images \citep{acc08},
were {\it a priori} optimized on data taken from the Crab nebula. An
event is considered to fall into the signal (ON) region once the squared
angular distance between the reconstructed event direction and the
Mrk\,421 position is $\Delta \theta^{2} \leq 0.025 \, \deg^{2}$. The
background is estimated from different regions of the same size
positioned at the same radial distance to the camera center as the ON
region, and is referred to as the reflected background region model
\citet{ber07}. The excess is then calculated as the number of ON source
counts less the normalized number of OFF region counts. In this analysis
5 OFF regions were used. The statistical significances are calculated
using the method of \citet{lima}.

The energy $E_{i}$ of an individual image is estimated using look-up
tables generated from {\it Monte Carlo} simulations of $\gamma$-ray air
showers. The tables are parameterized in (i) the integrated charge
(size), (ii) the impact parameter $p_{i}$ between the reconstructed
shower axis and the optical axis of telescope $i$, (iii) the zenith
angle $z$, (iv) the azimuth angle $Az$, and (v) the level of the
night-sky background. The energy of the shower event is then averaged
over the $n$ telescope energies to obtain $E_{\rm{reco}} = 1/n
\sum_{i=1}^{n} E_{i}$. The energy resolution is estimated based on {\it
Monte Carlo} simulations to be $\Delta E / E \approx 20\%$ for energies
between $100 \, \rm{GeV}$ and $30 \, \rm{TeV}$.

The effective areas $A_{\rm{eff}}$ describe the energy-dependent
response of the detector and are also obtained by {\it Monte Carlo}
simulations. The effective area $A_{\rm{eff}} (E_{\rm{reco}}, z, Az,
\Delta R, N_{\rm{tel}})$ is estimated for each event based on its
corresponding parameters; $\Delta R$ is the angular distance between the
reconstructed shower direction and the telescope pointing position,
$N_{\rm{tel}}$ is the number of telescopes in the system\footnote{A part
of the Mrk\,421 data were taken with only $N_{\rm{tel}} = 3$ operating
telescopes.}.

The inverse effective areas are used on an event-by-event basis to
calculate the differential photon flux for each bin $j$ in the energy
spectrum:

\begin{equation}
\label{formula:EnergySpectrumRezEffArea}
  \frac{\rm{d}N}{\rm{d}E_{j}} =
        \frac{1}{T^{\rm{live}}_{j} \cdot \Delta E_{j}}
                \left( \sum_{k=1}^{N_{\rm{on}, j}}
                        \frac{1}{A_{\rm{eff}, k}}
                - \alpha \sum_{k=1}^{N_{\rm{off}, j}}
                        \frac{1}{A_{\rm{eff}, k}} \right)
\end{equation}


The sum of $N_{\rm{on}}$ corresponds to reconstructed events from the ON
region and the sum of $N_{\rm{off}}$ to events from the five reflected
OFF regions with the normalization $\alpha = 0.2$. $T_{j}^{\rm{live}}$
is the live time for bin $j$. The bias energy describes the energy at
which the reconstructed energy $E_{\rm{reco}}$ (on average) deviates
less than a certain percentage from the true energy $E_{\rm{MC}}$. The
energy bias is calculated based on {\it Monte Carlo} simulations; a bias
threshold of $(E_{\rm{reco}} - E_{\rm{MC}}) / E_{\rm{MC}} = 10\%$ is
used in this analysis. The bias energy depends on the zenith angle of
the corresponding data run. For each $20 \, \rm{min}$ run, the bias
energy is calculated and only those bins, $j$, in the energy spectrum
which are fully contained above the bias energy are allowed to receive
events. The live time $T_{j}^{\rm{live}}$ is increased on a run-by-run
basis for only those energy bins.

In order to account for spill-over effects, the effective areas are 
calculated using the reconstructed energies, where the {\it Monte Carlo} 
input spectrum is weighted according to the reconstructed/measured 
spectrum in an iterative procedure. The systematic errors of the 
parameters describing a power-law energy spectrum $\rm{d}N/\rm{d}E = 
I_{0} \cdot E^{-\Gamma}$ have been estimated\footnote{Uncertainties in 
the atmosphere, components of the detector, and shower reconstruction 
algorithms were considered in this estimate.} based on a Crab-like 
energy spectrum ($\Gamma \approx 2.5$) to be: $\Delta \Gamma / \Gamma = 
8\%$ and $\Delta I_{0} / I_{0} = 20\%$.

The integral flux on a run-by-run basis above a certain energy $E'$ (as
shown in the light curves) is calculated as follows: A spectral slope is
assumed and the effective areas for the corresponding run parameters
(zenith angle, etc.) are used together with the measured excess to
determine the normalization. The normalization is then used to calculate
the integral flux above $E'$. This procedure has the advantage that the
full event statistics are used and $E'$ is not limited by the strongly
varying thresholds of individual data subsets (i.e. runs taken at
different zenith angles). However, a spectral shape has to be assumed,
which in our case is chosen (iteratively) for each data point according
to the energy spectrum corresponding to the estimated flux level.

Mrk\,421 is one of the objects in a trigger agreement between the
$\gamma$-ray observatories H.E.S.S., MAGIC, and VERITAS put into place
for known TeV $\gamma$-ray blazars in order to exchange information
about flaring sources. The trigger criterion for Mrk\,421 is defined by
a flux level measured by one of the observatories exceeding a value of
$2 \Phi_{\rm{crab}}$; this criterion was met several times during the
campaign in 2008, leading to triggers sent by VERITAS and to triggers
received from the MAGIC collaboration. VERITAS observed Mrk\,421 during
January/February and November/December 2007 ($5 \, \rm{h}$) as well as
in January-June 2008 ($42.3 \, \rm{h}$) for a total of $47.3 \, \rm{h}$
after run quality selection.  The observation time corrected for the
detector dead time amounts to $43.6 \, \rm{h}$. The zenith-angle range
of the observations was $6 \deg - 56 \deg$ with an average of $23.5
\deg$, corresponding to an analysis energy threshold\footnote{The energy
threshold is defined as the energy corresponding to the peak detection
rate for a Crab-like spectrum.} of $260 \, \rm{GeV}$.

\paragraph{Whipple.}

The $10 \, \rm{m}$ $\gamma$-ray Telescope at the Fred Lawrence Whipple
Observatory \citep{Kil07} is sensitive in the energy range from $200 \,
\rm{GeV}$ to $20 \, \rm{TeV}$ with a peak response energy (for a
Crab-like spectrum) of approximately $400 \, \rm{GeV}$. This telescope,
although a factor of seven less sensitive than VERITAS, was used in this
program to extend the TeV coverage when VERITAS was not available for
Mrk 421 observations. More detailed descriptions of Whipple observing
modes and analysis procedures can be found elsewhere
\citep{weekes,Punch1991,reynolds}. Details about the Whipple telescope
including the GRANITE-III camera have been given in \citet{Kil07}.

The Whipple observations were conducted between November 2005 and May
2008 (MJD 54417--54622). Only runs which pass the run quality selection
(stability of the raw trigger rate, induced by the cosmic ray
background) are considered in the analysis resulting in a data set of
$96 \, \rm{h}$ of ON-source data. The data were analyzed using the
standard $2^{\rm{nd}}$-moment-parameterization technique
\citep{Hill:85}.  Standard cuts ($\verb SuperCuts 2000$) were used to
select $\gamma$-ray events and to suppress background cosmic-ray events
\citep{Call:03}. Using the zenith angle dependence of a Crab data set
taken at similar epochs, we account for the zenith angle dependence of
the $\gamma$-ray excess rate by normalizing our measured Mrk\,421 rate
to the Crab rate at a corresponding zenith angle. It should be noted
that this simplistic scaling is strictly only valid for a TeV spectrum
close to that of the Crab Nebula (spectral index of $\Gamma = 2.5$).
However, the systematic error introduced by this scaling can be expected
to be small compared to the statistical error of the flux points.

\subsection{{\it RXTE, SWIFT} and {\it Suzaku} X-ray Observations}
\label{subsec:DataXRay}
 
X-ray data were taken with the telescopes on the Rossi X-ray Timing
Explorer {\it RXTE} \citep{Swank1994}, {\it Swift} \citep{Gehrels2004},
and {\it Suzaku} \citep{Mitsuda2007} satellites. The pointed X-ray
observations are summarized in Table~\ref{tab:ObservationsXRay}.

\paragraph{RXTE/PCA.}

The Proportional Counter Array (PCA) \citep{Jahoda1996} comprises 5
Proportional Counter Units (PCUs) covering a nominal energy range of
$2-60 \, \rm{keV}$ with a net detection area of $6250 \, \rm{cm}^{2}$. 
Data between the energies of $3 - $15 \, \rm{keV} were used in this
analysis. The $15-250 \, \rm{keV}$ data from the High-Energy X-ray
Timing Experiment {\it HEXTE} \citep{Rothschild1998} were not used owing
to an insufficient signal-to-noise ratio. The PCA data were taken as
part of a multi-wavelength observation proposal and comprise 161
exposures between January 2006 and May 2008 (see
Table~\ref{tab:ObservationsXRay}) with a total net exposure time of
$245.6 \, \rm{ksec}$. The observations had a typical exposure of $10-70
\, \rm{min}$ per pointing and were taken at near-simultaneous times to
scheduled VERITAS observations\footnote{The RXTE/PCA staff at NASA GSFC
and Principal Investigator (PI) of the observation proposal Henric
Krawczynski, together with the VERITAS team, coordinated the
observations.}. For the observations from January~6, 2006 to April~18,
2006 both PCU0 and PCU2 detectors collected data, while for all other
data only PCU2 was operational. The data were filtered following the
standard criteria advised by the NASA {\it Guest Observer Facility
(GOF)}\footnote{http://heasarc.gsfc.nasa.gov/docs/xte/xte\_1st.html}. 
Standard-2 mode PCA data gathered with the top layer (X1L and X1R) of
the operational PCUs were analyzed using the HEAsoft 6.4 package. 
Background models were generated with the tool {\it pcarsp}, based on
the {\it RXTE} GOF calibration files for a ``bright'' source with more
than 40 counts/sec.  Response matrices for the PCA data were created
with the script {\it pcarsp}. The {\it saextrct} tool was used to
extract all PCA energy spectra.

\paragraph{RXTE/ASM.}

The RXTE All Sky Monitor (ASM) \citep{Levine1996} is sensitive to X-ray
energies at $2-12 \, \rm{keV}$ and scans most of the sky every 1.5
hours. The data were obtained from the public MIT
archive\footnote{http://heasarc.gsfc.nasa.gov/docs/xte/asm\_products.html}
in the form of 1-day averaged binning, as well as the dwell-by-dwell
binning (for the short-term light curve and flux correlation studies).

\paragraph{Swift/XRT.}

The X-ray telescope (XRT) on board the Swift satellite
\citep{Gehrels2004} is a focusing X-ray telescope with a $110 \,
\rm{cm}^{2}$ effective area and a $23 \arcmin$ field of view
\citep{Burrows2005}. It is sensitive to X-rays in the $0.2 - 10 \,
\rm{keV}$ band. A total of $1175.7 \, \rm{ksec}$ of XRT data were taken
between January 2006 and May 2008 (Table~\ref{tab:ObservationsXRay})
in the Windowed Timing (WT) mode with grades 0-2 (referring to the
pattern of CCD pixels for each event) selected over the energy range
$0.4 - 10 \, \rm{keV}$.  The XRTPIPELINE tool was used to calibrate and
clean all Swift XRT event files with current calibration files. The data
were reduced using the HEAsoft~6.4 package. Source counts were extracted
from a rectangular region of 40 pixels (94.4 arcsec) along the one
dimensional stream, and 20 pixels high centered on the source.
Background counts were extracted from a nearby source-free rectangular
region of equivalent size. Ancillary response files were generated using
the {\it xrtmkarf} task applying corrections for the PSF losses and CCD
defects. The latest response matrix from the XRT calibration files was
used. The extracted XRT energy spectra were re-binned to contain a
minimum of 20 counts in each bin and were fit with XSPEC 12.4.

\paragraph{Swift/BAT.}

The Burst Alert Telescope (BAT) is a large field of view (1.4 steradians)
X-ray telescope with imaging capabilities in the energy range from $15 -
150 \, \rm{keV}$ \citep{Gehrels2004}. The BAT typically observes 50\%
to 80\% of the sky each day. The data are the Swift/BAT transient
monitor results provided by the Swift/BAT team \citep{Krimm2008a}. Full
details of the BAT data analysis are given at the BAT transients web page
\citep{Krimm2008b}.

\paragraph{Suzaku/XIS.}

The X-ray Imaging Spectrometer (XIS) \citep{KO07} on board the Suzaku
satellite is composed of 3 X-ray CCD cameras combined with a single
X-ray Telescope (XRT) covering a nominal energy range of $0.5 - 12 \,
\rm{keV}$. Each CCD camera covers an $18 \arcmin \times 18 \arcmin$
region of the sky. The XIS data include observations between May 5, 2008
and May 8, 2008 with an exposure time of $180.8 \, \rm{ksec}$. Standard
data reduction and processing were performed using HEAsoft v6.6.3 and
ftools v6.6. XIS events were extracted from a source region with an
inner radius of 35 pixels and an outer radius of 408 pixels. The extent
of the inner radius is such that pile-up effects were minimized for the
selected events. The background was selected from an annulus outside of
the source region defined by 432 pixel and 464 pixel inner and outer
radii respectively. The response matrix and effective area were
calculated for each XIS sensor using the Suzaku ftools tasks,
\textit{xisrmfgen} and \textit{xissimarfgen} \citep{IS07}. XIS1 data
were not included in this analysis. As the XIS0 and XIS3 have similar
responses, their data were summed.

\begin{table}[t]
\begin{center}

\caption{\label{tab:ObservationsXRay} Pointed X-ray observations 
(RXTE/PCA, Swift/XRT, and Suzaku/XIS) of Mrk\,421 in 2006 to 2008.}

\begin{tabular}{lrrr}

\noalign{\smallskip}
\tableline\tableline
 Start & Stop & $N_{\rm{obs}}$ & ObsID \\

\noalign{\smallskip}
\tableline\tableline

\noalign{\smallskip}

\multicolumn{4}{l}{RXTE/PCA} \\
\tableline
\noalign{\smallskip}

 2006-01-06 & 2006-03-02 & 27 & 91440-01 \\
 2006-03-03 & 2006-05-31 & 48 & 92402-01 \\
 2008-01-07 & 2008-05-07 & 86 & 93133-02 \\

\noalign{\smallskip}

\multicolumn{4}{l}{Swift/XRT} \\
\tableline
\noalign{\smallskip}

 2006-01-02 & 2006-12-05 & 20 \\
 2007-03-23 & 2007-12-31 & 24 \\
 2008-01-07 & 2008-05-08 & 54 \\

\noalign{\smallskip}

\multicolumn{4}{l}{Suzaku/XIS} \\
\tableline
\noalign{\smallskip}

 2008-05-05 & 2008-05-08 & 1 \\

\tableline
\end{tabular}

\end{center}
\end{table}

\subsection{Optical Observations}
\label{subsec:DataOptical}

Many optical observatories contributed data sets to this campaign (see
below). The data from the observatories were reduced and the photometry
performed independently by different analysts using different
strategies. The same set of reference stars was used for all optical
data sets to calculate the systematic error on the flux. However,
combining the various optical data to produce a composite light curve
for each spectral band is complicated by the fact that different
observatories use different photometric systems.  Furthermore,
photometric apertures and the definition of the reported measurement
error for each nightly-averaged flux is inconsistent across datasets. 
Therefore we have adopted a simple approach for the construction of the
composite light curves whereby a unique flux offset is found for each
spectral band (R, B, V) of every instrument based on overlapping
observations \citep{Ste07}, and the light curves have been scaled
accordingly (in our case the light curves of the Bradford Robotic
Telescope and the New Mexico Skies observatory by $15\%$ each).

\paragraph{UVOT.}

Mrk\,421 was observed with the Swift Ultraviolet/Optical Telescope
(UVOT) during 2008. The instrument cycled through each of three
ultraviolet pass bands, UVW1, UVM2 and UVW2 with central wavelengths of
$260 \rm{nm}$, $220 \rm{nm}$ and $193 \rm{nm}$, respectively. More than
100 observations were obtained with a typical/average exposure (per
filter) of $150 \, \rm{s}$, ranging from $50 \, \rm{s}$ up to $900 \,
\rm{s}$. Data were taken in the {\it image mode}, where the image is
accumulated on board the satellite discarding the photon timing
information within each single exposure to reduce the telemetry volume
and the time of transmission.

Primary and secondary analyses were carried out using UVOTSOURCE
standard tool and a custom UVOT pipeline. Both analyses used the
calibration database released on February 2010. Photometry was computed
using a $5 \, \arcsec$ source region around the source and photometric
corrections were applied following \citet{Poole2008} and \citet{Li2006}.
All observations were inspected manually. Astrometric misalignment
between the observed position and the nominal position of Mrk\,421 which
were found in several data sets were corrected by using a spatial
fitting algorithm. Results of the two analysis chains were found to be
in agreement.

Due to the ultraviolet spectrum, we adopted the rate-to-flux conversion
factors for GRB-like objects and not the standard factors used for
Pickles-like star spectra. The fluxes were corrected for galactic
extinction $E_{B-V} = 0.015 \, \rm{mag}$ \citep{Schlegel1998}. To obtain
this value we smoothed the nearer values of the database in
correspondence with the coordinates of the source. Then, the computed
optical/UV galactic extinction coefficients were applied
\citep{FitzMassa1999}. The effects of intergalactic absorption and the
zodiacal light have been estimated to be negligible and were not
corrected for in this analysis. The fluxes and corresponding frequencies
shown in the light curves are redshift corrected, including a
second-order correction taking into account filter non-linearities.

The host galaxy correction was not applied, but a systematic error on
the flux is estimated. The measurements of \citet{Nilsson1999} are used
to estimate the host galaxy emission in the $R$ band. These are used to
obtain the corresponding components for the $V$, $B$ and $U$ bands
\citep{Fukugita1995}. The presence of an upturn flux excess in the far UV
spectrum of elliptical galaxies is caused by an old population of hot
helium-burning stars without extended hydrogen-rich envelopes (as
compared to rather young stars). The findings of \citet{Arimoto1996}
were used to calculate the metallicity of the Mrk\,421 host galaxy and
therefore constrain its contribution to the UV bands to be less than
$5\%$ \citep{Han2007}.

Some caveats have to be mentioned. There are bright sources in the field
of view which will cause significant coincidence losses, ghosting from
internal reflections, and may lead to an additional overestimation of
the Mrk\,421 blazar flux. Although the relative photometry (light
curves) is expected to be less sensitive to these effects, an additional
systematic error of $20 \%$ was added to the absolute UVOT fluxes shown
in the SEDs to account for these uncertainties.

\paragraph{BRT/NMS.}

Optical data were taken with the Bradford Robotic Telescope (BRT) in
Tenerife, Canary Islands, Spain, as well as the New Mexico Skies
observatory (NMS). The data were reduced by standard aperture
photometry\footnote{The MIRA Pro Version~7 (Mirametrics, Inc.) was used,
see \url{http://www.mirametrics.com/}.}. The aperture size used was $10
\arcsec$ diameter, and the comparison stars were taken from Villata et
al. (1998).

\paragraph{Bell.}

The Western Kentucky University's Bell observatory is a $0.6 \, \rm{m}$
telescope located 12 miles southwest of Bowling Green, Kentucky. The
observations presented here were obtained with an AP6 CCD camera and
Bessell R band filter. Dark and flat field corrections were made to the
images and differential aperture CCD photometry was performed using
stars 1,3,2 of the comparison sequence from Villata et al. (1998). No
correction for host galaxy flux or galactic absorption was made.

\paragraph{WIYN.}

The WIYN $0.9 \rm{m}$ telescope is located at the National Optical
Astronomy Observatory at Kitt Peak and operated by a consortium of
universities. Observations of Mrk\,421 were performed since January 2006
in Johnson B and V and Cousins R optical filters and a CCD $1 \deg$
field of view Mosaic Imager. Image reduction was performed with IRAF,
using bias frames and dome flat fields for each night of data. We
obtained magnitudes by differential photometry, using three reference
stars from Villata et al. (1998). Since we are mostly interested in
measuring the magnitude relative variations with time, these data were
not corrected for the host galaxy flux or absorption.

\paragraph{Tuorla/KVA.}

The Kungliga Vetenskapsaka-demien telescope (KVA, Royal Swedish Academy
of Sciences) is located on Roque de los Muchachos, La Palma and operated
by the Tuorla Observatory, Finland.  The telescope is composed of a $0.6
\, \rm{m}$ f/15 Cassegrain devoted to polarimetry, and a $0.35 \,
\rm{m}$ f/11 SCT auxiliary telescope for multicolor photometry.  This
telescope has been successfully operated in a remote way since autumn
2003. Mrk\,421 has been observed in optical R-band typically once per
night.  Photometric measurements were made in differential mode, i.e. by
obtaining CCD images of the target and calibrated comparison stars in
the same field of view \citep{FiorucciTosti1996, Fiorucci1998,
Villata1998}.


\subsection{Radio Observations \label{subsec:DataRadio}}

Radio data presented here were taken at four frequencies at two
different radio observatories. The fluxes are given in Janskys (Jy), so
they have already been normalized for the bandwidth of their receivers.

\paragraph{Mets\"ahovi}

The Mets\"ahovi radio telescope (radome enclosed paraboloid antenna,
diameter of $13.7 \, \rm{m}$) is situated in Finland. The measurements
were made with a $1 \, \rm{GHz}$-band dual beam receiver centered at
$36.8 \, \rm{GHz}$. The observations are ON-ON observations (typical
integration time of $1200-1400 \, \rm{s}$), alternating the source and
the sky in each feed horn.

The detection limit of the telescope at $37 \, \rm{GHz}$ is of the order
of $0.2 \, \rm{Jy}$ under optimal conditions. Data points with a
signal-to-noise ratio $< 4$ are handled as non-detections. The flux
density scale is set by observations of DR 21. Sources 3C 84 and 3C274
are used as secondary calibrators. A detailed description of the data
reduction and analysis is given in \citet{Ter98}. The error estimate in
the flux density includes the contribution from the measurement rms and
the uncertainty of the absolute calibration.

\paragraph{UMRAO.}

The University of Michigan Radio Astronomy Observatory (UMRAO) ($26 \,
\rm{m}$ paraboloid) provided monitoring data of Mrk\,421 at $4.8 \,
\rm{GHz}$, $8 \, \rm{GHz}$ and $14.5 \, \rm{GHz}$ between June 2006 and
May 2008. Each observation consisted of a series of ON-OFF measurements
taken over a 30-40 minute time period. All observations were made within
a total hour angle range of about 5 hours centered on the meridian.  The
calibration and reduction procedures have been described in
\citet{Aller1985}. Some daily observations were averaged to improve the
signal-to-noise ratio.

Unfortunately, the source is rather weak which may mask some
variability. Nevertheless, the UMRAO measurements over many decades have
identified continuous fluctuations in amplitude but not a single
outburst-like flare. Small structural changes in the radio jet are
apparent in the MOJAVE $15 \, \rm{GHz}$ VLBI images for the
source\footnote{Mrk\,421 on the MOJAVE (monitoring of jets in active
galactic nuclei with VLBA experiments) project page:
http://www.physics.purdue.edu/astro/MOJAVE/sourcepages/1101+384.shtml}.

\section{Results \label{sec:results}}

In the whole VERITAS data set an excess of $29974$ $\gamma$-ray events
was detected from the direction of Mrk\,421 after application of event
selection cuts ($31523$ ON events, $7746$ OFF events, normalization
$\alpha = 0.2$), corresponding to a statistical significance of
$277$~standard deviations. An overview of the light curves at radio to
TeV energies is given in Section~\ref{subsec:ResultsOverview}.
Subsequently, we discuss the time and spectral variability of the TeV
$\gamma$-ray data (Section~\ref{subsec:ResultsTeV}) as well as the time
and spectral variability of the X-ray fluxes
(Section~\ref{subsec:Results_XRay}) on different time scales. Finally,
we scrutinize how those relate to the flux and spectral variability in
other energy bands in Section~\ref{subsec:ResultsCross}.

\subsection{Light Curves\label{subsec:ResultsOverview}}

\begin{figure*}[p]

\centering{
\includegraphics[width=0.88\textwidth]{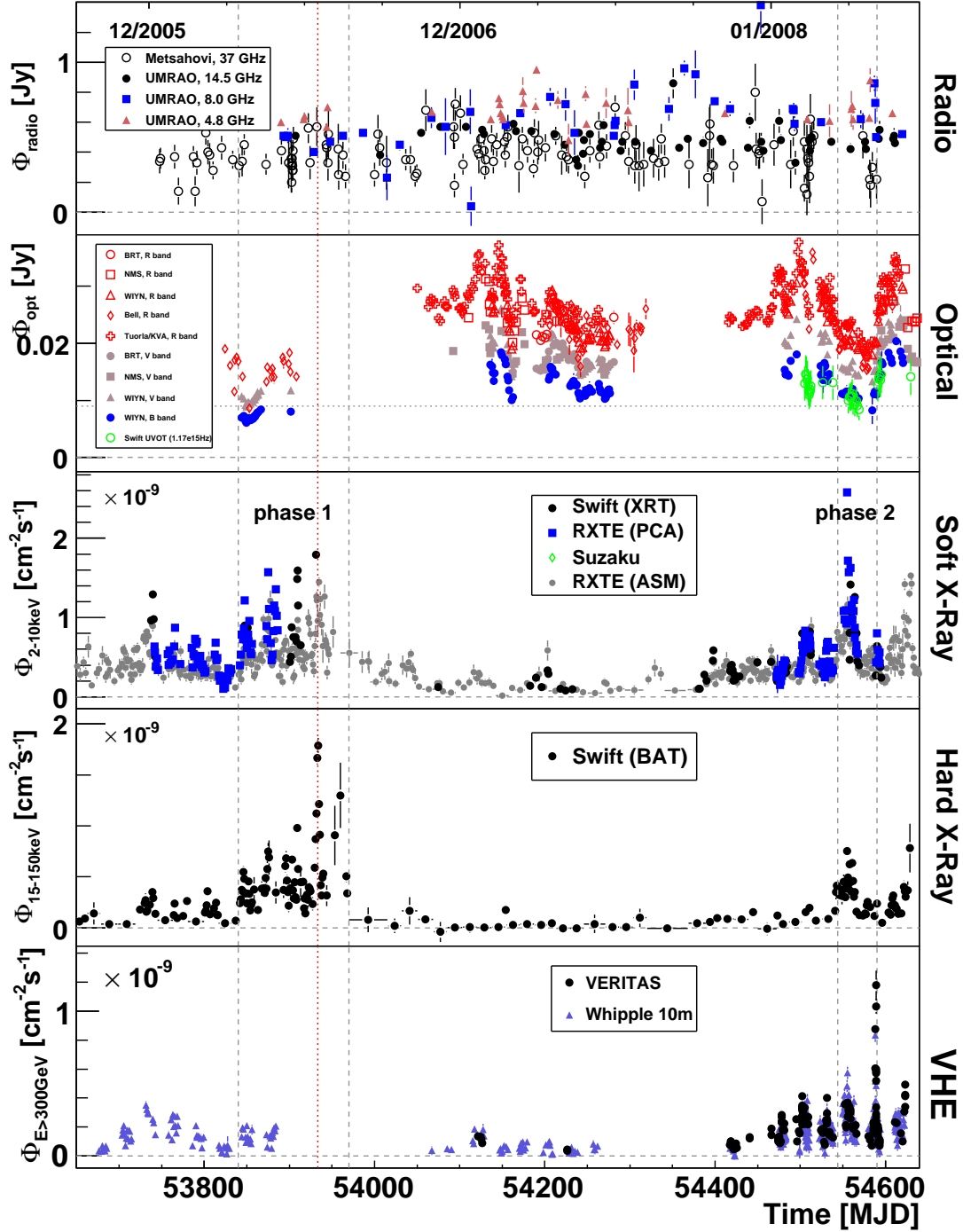}
}

\caption{\label{fig:MWL_LightCurveAll} Mrk\,421 light curves measured by
different experiments (see Section~\ref{sec:datasets}) in 2006-2008. All
errors are statistical errors only. Shown are the radio band (upper
panel), the optical band (second panel), the X-ray band (third \& fourth
panel) and the TeV band (bottom). Some of the RXTE/ASM and Swift/BAT
flux points have been re-binned for better visibility. Two phases of
activity can be identified in the X-ray and TeV bands (phases 1 and 2). 
The vertical dotted line indicates the maximum of the hard X-ray flux
(Swift/BAT, phase~1). During phase~2 a very good X-ray/TeV coverage was
achieved, where the X-ray observations were partly triggered by
VERITAS.}

\end{figure*}

\begin{figure*}[p]

\centering{
\includegraphics[width=0.88\textwidth]{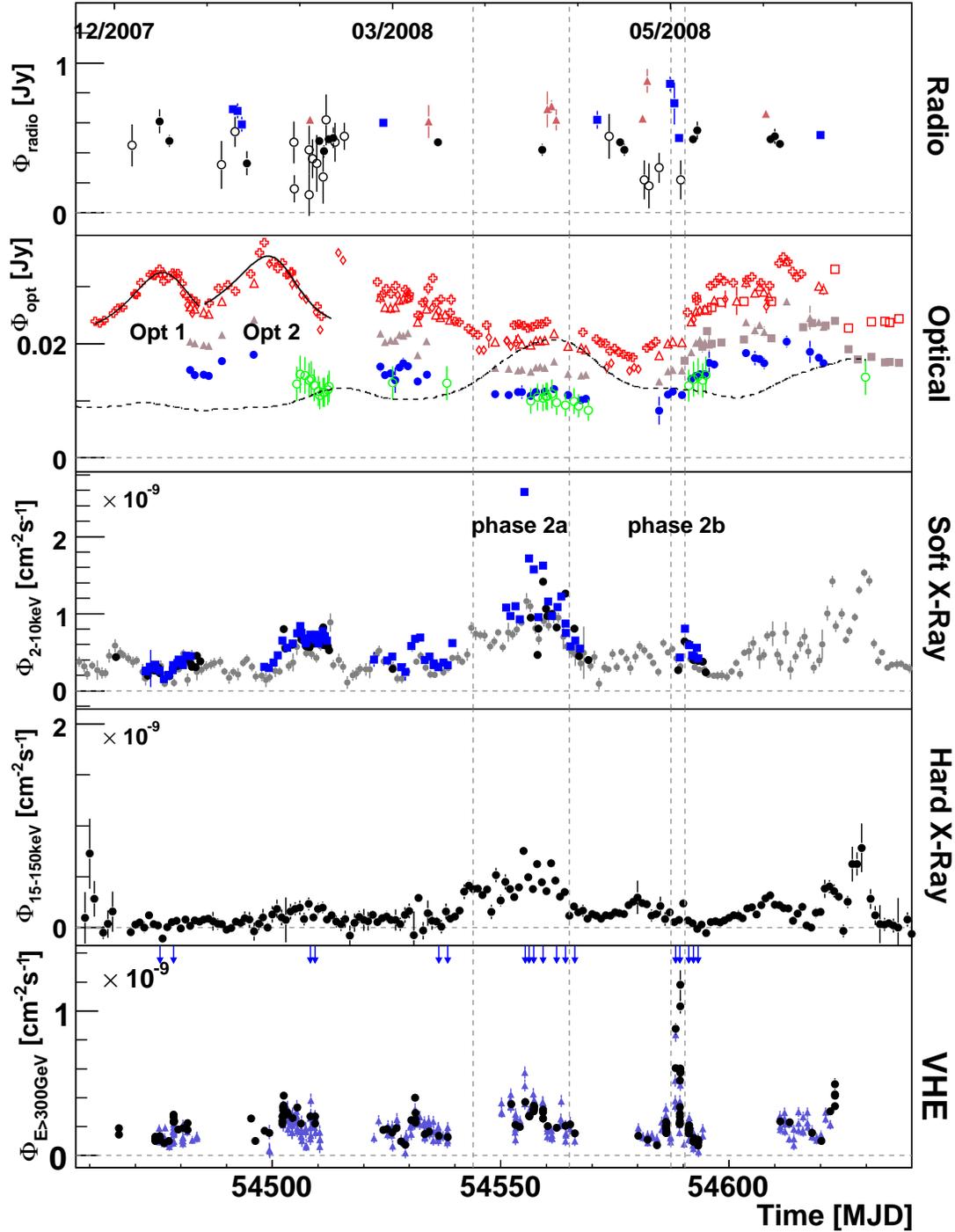}
}

\caption{\label{fig:MWL_LightCurveZoom} 2008 Mrk\,421 light curves. For 
details see caption and legends in Figure~\ref{fig:MWL_LightCurveAll}. 
The vertical lines enclose phases of enhanced X-ray (phase~2a) and 
enhanced TeV (phase~2b) activity; see 
Figure~\ref{fig:MWL_LightCurveZoom2} for a zoomed view of these phases. 
An exponential rise/fall function was fitted to the optical R band 
Tuorla/KVA data ('Opt\,1' \& 'Opt\,2', see discussion in 
Sec.~\ref{subsec:ResultsCross} and fit results in 
Tab.~\ref{tab:OptFlares}). The dashed line represents the optical 
rise/fall function folded with the ASM X-ray (dwell-by-dwell) light 
curve (see text for disussion). The vertical arrows pointing downwards 
in the bottom panel indicate the dates for which we compiled 
quasi-simultaneous SEDs and modeled them with a SSC code (see 
Figures~\ref{fig:SEDs1} \& \ref{fig:SEDs2}).}

\end{figure*}

\begin{figure*}[p]

\centering{
\includegraphics[width=0.88\textwidth]{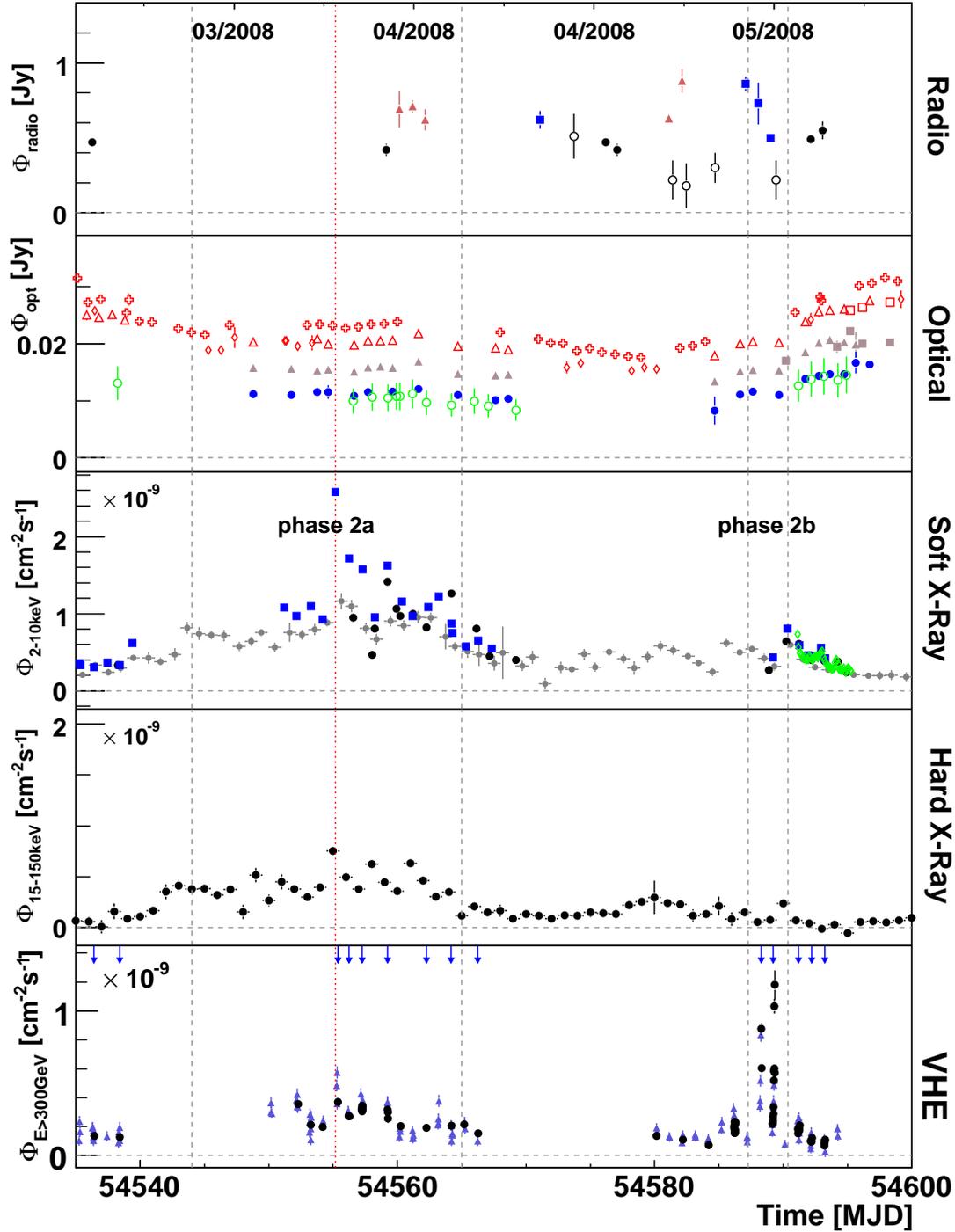}
}

\caption{\label{fig:MWL_LightCurveZoom2} Mrk\,421 light curves measured
during the X-ray/TeV flaring state in April/May 2008 (phase~2). For
details see caption and legends in Figure~\ref{fig:MWL_LightCurveAll}. 
The vertical dotted line highlights the brightest X-ray flare observed
with the RXTE/PCA. The vertical arrows pointing downwards in the bottom
panel indicate the dates for which we compiled quasi-simultaneous SEDs
and modeled them with a SSC code (Figures~\ref{fig:SEDs1} \&
\ref{fig:SEDs2}).}

\end{figure*}

\begin{figure*}[p]

\centering{
\includegraphics[width=0.88\textwidth]{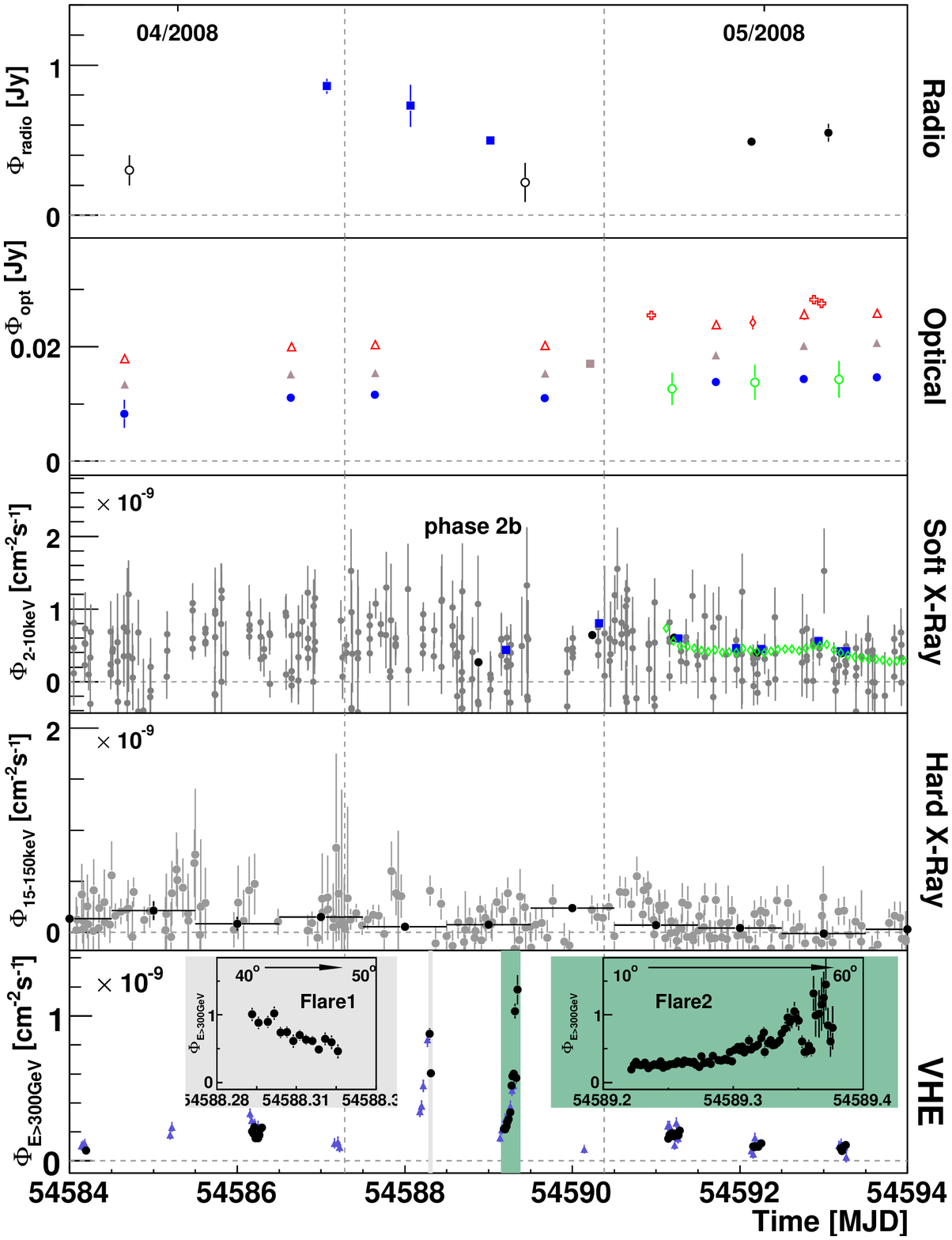}
}

\caption{\label{fig:MWL_LightCurveZoom3} Mrk\,421 light curves measured
during the TeV flaring state in April/May 2008 (phase~2b). For details
see caption and legends in Figure~\ref{fig:MWL_LightCurveAll}. Note,
that the highest TeV flux around MJD~54590 was not accompanied by an
increase in X-ray activity (compare with
Figure~\ref{fig:Corr_VHE_Xray}). The continuous Suzaku/XIS soft X-ray
monitoring (starting two days after the strong TeV flare) is shown, as
well. The inserts in the lowest panel show the VERITAS TeV light curves
in a $3 \, \rm{min}$ binning, covering the flaring time span indicated
by the colored, vertical boxes. The flares were recorded under mostly
large zenith angles (indicated by the arrows) with a correspondingly
higher systematic error on the flux, see text for more details.}

\end{figure*}

The radio, optical, X-ray and TeV light curves of Mrk\,421 are shown in 
Figure~\ref{fig:MWL_LightCurveAll} for the years 2006-2008 during which 
the source was extensively monitored in the various energy bands. A 
zoomed version for the year 2008 is shown in 
Figure~\ref{fig:MWL_LightCurveZoom}. Two further levels of zoom are shown 
in Figure~\ref{fig:MWL_LightCurveZoom2} (the two more active X-ray/TeV 
states) and Figure~\ref{fig:MWL_LightCurveZoom3} (the strong TeV 
$\gamma$-ray flare).

For a clearer representation, the RXTE/ASM and Swift/BAT night-by-night
data points in Figures~\ref{fig:MWL_LightCurveAll} to
\ref{fig:MWL_LightCurveZoom2} were combined until one of the following
conditions was met: (i) the combined data point had a statistical
significance of more than $3$~standard deviations ($\sigma$), or (ii) 15
bins of the original light curve were combined. The RXTE/ASM and
Swift/BAT data in Figure~\ref{fig:MWL_LightCurveZoom3} are shown in
dwell-by-dwell bins. The TeV data (Whipple and VERITAS) in all figures
are shown in a run-by-run binning of $10-20 \, \rm{min}$ duration. All
other data are shown in a binning corresponding to the
pointings/exposures of the individual experiments.

Although the X-ray fluxes show a long-term structure with phases of 
higher activity followed by phases of lower activity, flux variations by 
a factor of two can be observed on time scales down to a few days. In 
the optical band, flux variations are observed on longer time scales of 
the order of a weeks to months. The structure of the optical light curve 
(i.e. Figure~\ref{fig:MWL_LightCurveZoom} and possible connections to 
the X-ray/TeV band are discussed in Sec.~\ref{subsec:ResultsCross}. No 
significant flux variations can be observed at radio energies. The flux 
correlations between the different energy bands are discussed in 
Section~\ref{subsec:ResultsCross}. Two phases of enhanced X-ray and/or 
TeV activity can be identified in the light curve:

\begin{itemize}

\item {\bf Phase~1:} The first active phase ({\it phase~1}) occurred
during the summer 2006 and lasted for at least a few months. The active
state can be identified in soft and hard X-rays
(Figure~\ref{fig:MWL_LightCurveAll}). During this time period no VERITAS
data were taken and only a few nights are covered by Whipple data.

\item {\bf Phase~2} The second active phase ({\it phase~2}) occurred in
April/May 2008 and was recorded with excellent coverage in the X-ray and
TeV bands (Figure~\ref{fig:MWL_LightCurveZoom}). However, a zoom-in of
this second flaring phase (Figures~\ref{fig:MWL_LightCurveZoom2} and
\ref{fig:MWL_LightCurveZoom3}) shows that the strongest TeV emission
({\it phase~2b}) is not coincident with the strongest soft/hard X-ray
activity ({\it phase~2a}, peaking roughly one month before phase~2b).
The lack of increased X-ray emission during the peak TeV flaring might
indicate an orphan flare \citep{Krawczynski_1ES1959}. However, the
characteristic time scales of flux changes in the TeV band can be less
than an hour (the major flare is fully contained within a time interval
of $5 \, \rm{h}$), so that a detailed comparison has to be restricted to
closely-simultaneous data, see Section~\ref{subsec:ResultsCross}. The
TeV flare is followed by a somewhat enhanced X-ray flux: The Swift/BAT,
Swift/XRT and RXTE/PCA data indicate a doubling in flux level between
the night of the flare and the following night, declining back to the
previous level within a few days, which is nicely sampled by the
Suzaku/XIS (Figure~\ref{fig:MWL_LightCurveZoom3}). The corresponding
structure of the X-ray light curve, however, does not substantially
differ from low-state variations, so that a physical connection to the
TeV activity cannot be claimed.

\end{itemize}

\subsection{Temporal and Spectral Variability in the TeV $\gamma$-Ray 
Band \label{subsec:ResultsTeV}}

\begin{figure*}[t]

\includegraphics[width=0.9\textwidth]{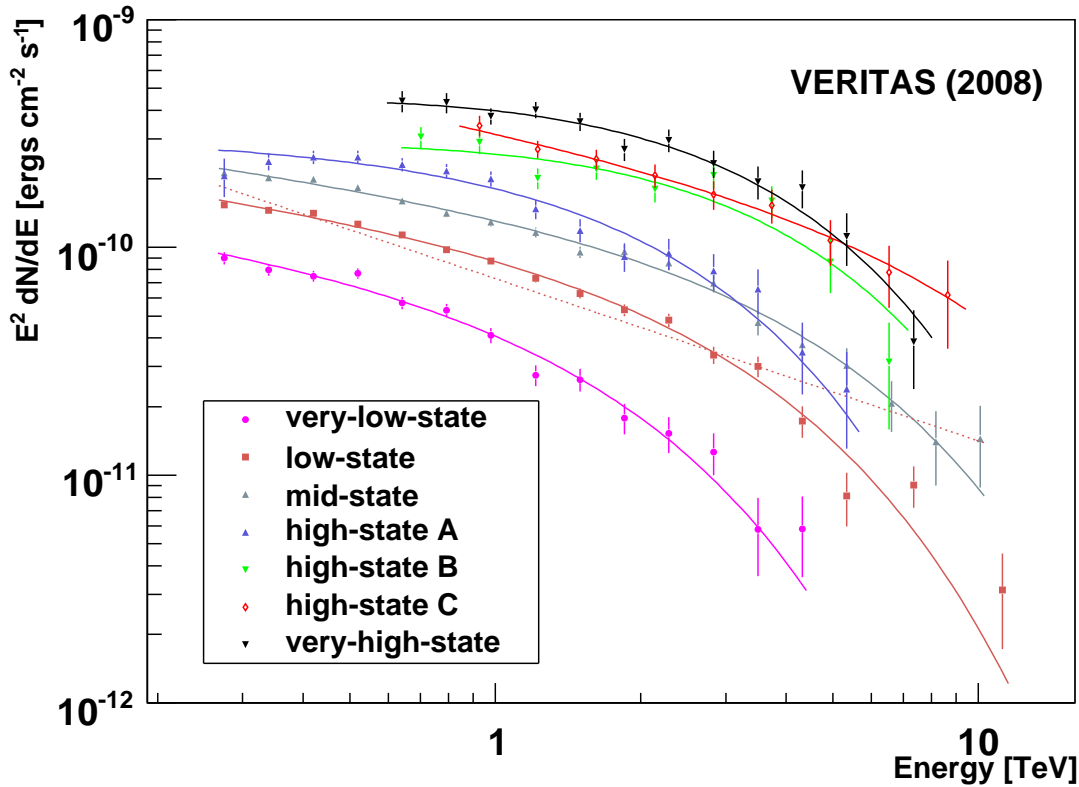}

\caption{\label{fig:VER_SpecFluxLevels} Time-averaged VERITAS energy
spectra for different flux levels. Only data taken in 2008 are used. A
power-law with exponential cutoff is fit to each spectrum (curves), the
fit parameters and statistical errors are summarized in
Table~\ref{tab:SpectrumTeV}. The dotted line shows a power-law fit to
the low-state spectrum to illustrate the incompatibility of this model
and the data. The high-flux spectra were taken at zenith angles $z>35
\deg$ with a correspondingly higher energy threshold.}

\end{figure*}

\paragraph{Flux variability.}

Flux variations on time scales of 1-2 days are found in the TeV band. 
Except for two nights measured during a strong flare in May 2008
(phase~2b, Figure~\ref{fig:MWL_LightCurveZoom2}), no significant TeV
flux variations are observed within individual nights. It should however
be mentioned, that the observation time during individual nights often
did not exceed $10-20 \, \rm{min}$\footnote{Regular run durations are
$20 \, \rm{min}$, but for monitoring purposes, some observations were
conducted with just $10 \, \rm{min}$ per night.}. This prevents placing
strong constraints on the $\gtrsim$0.5$-$1~h time scale variability,
given the low/medium states of Mrk\,421 during most of the measurements.
Nevertheless, the strong outburst measured in May 2008
(Figure~\ref{fig:MWL_LightCurveZoom3}) clearly shows variability on
sub-hour time scales even though the flare was recorded at zenith angles
down to $60 \, \deg$ (see inserts in
Figure~\ref{fig:MWL_LightCurveZoom3}). At high zenith angles $z$, the
effective areas vary drastically with a small change in $z$ and the
sensitivity suffers from the strongly increased energy threshold
(leading to a huge loss in event statistics). Since the data points in
the light curve are given above $300 \, \rm{GeV}$, the calculated fluxes
(for the zenith angle range of $40\deg - 60\deg$) are derived
(extrapolated) based on the measured spectral shapes ($\rm{d}N/\rm{d}E =
I_{0} \cdot (E/1\,\rm{TeV})^{-\Gamma} \cdot \exp(-E/E_{\rm{cut}})$).
Therefore, an increased systematic error on the integral flux of
$\sim$30$\%$ for $z> 45\deg$ and $\sim$40$\%$ for $z > 55\deg$ is
assumed. Given these facts, we do not determine a quantitative value for
the flux doubling times observed during this flare.

\paragraph{Spectral variability.}

\begin{figure}[t]

\epsscale{0.99}
\plotone{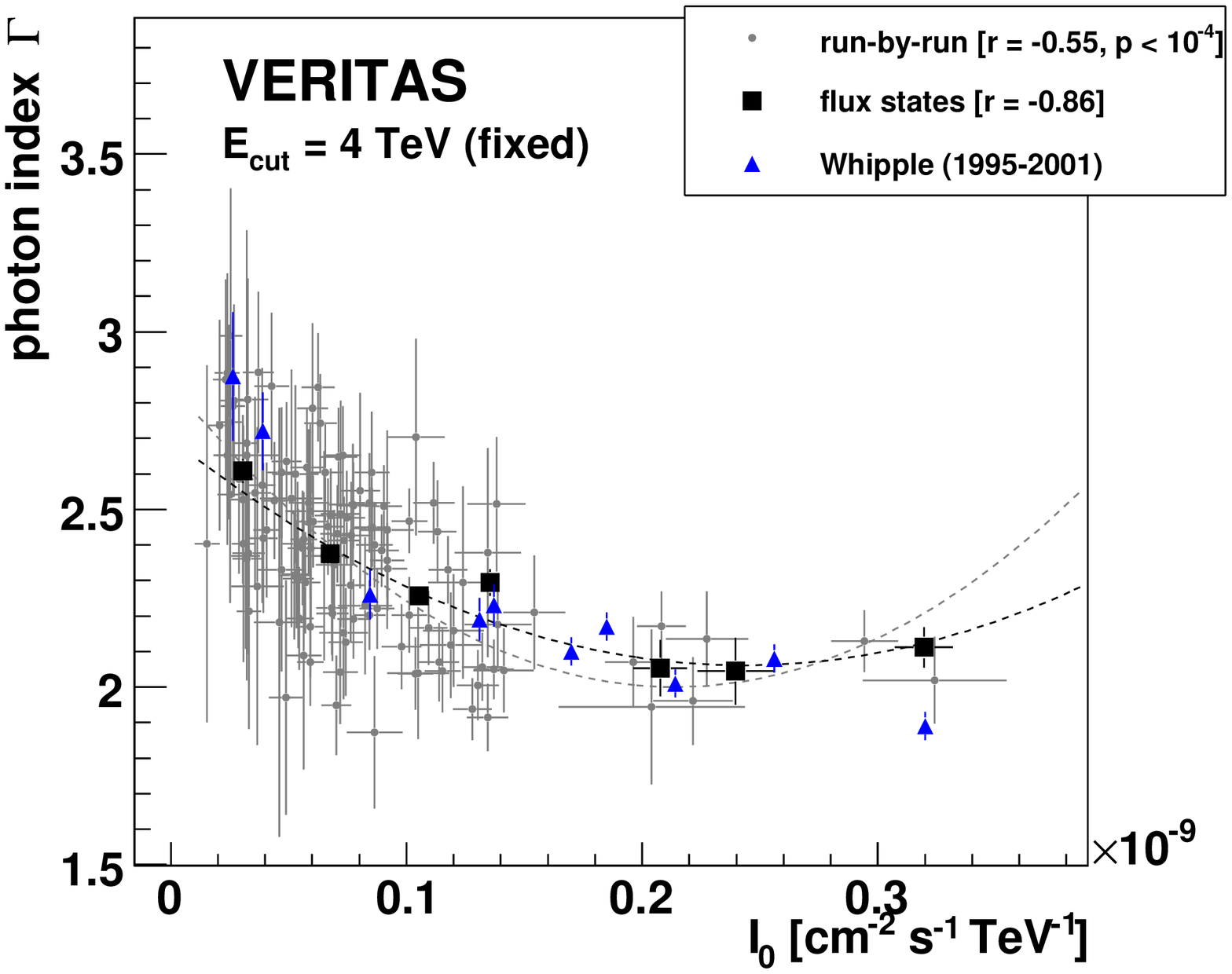}

\caption{\label{fig:VHE_FluxVsIndex} Photon index $\Gamma$ versus flux
normalization $I_{0}$ (at $1 \, \rm{TeV}$) obtained from a power-law fit
with exponential cutoff (fixed to $E_{\rm{cut}} = 4 \, \rm{TeV}$,
statistical errors only). Each gray data point corresponds to a data run
with a duration of $10$ to $20 \, \rm{min}$. The black points correspond
to the values obtained from the fits to the spectra grouped according to
their flux states (defined in Figure~\ref{fig:VER_SpecFluxLevels} and
Table~\ref{tab:SpectrumTeV}). Correlation coefficients $r$ and the
corresponding chance probabilities for the null hypothesis (no
correlation) are given in the legend. Results obtained from earlier
Whipple 10m observations \citep{Krennrich2002} are shown for reference.}

\end{figure}

To investigate a possible change of the spectral shape as a function of
flux state, the data are divided into subsets according to different
flux levels. The separation into flux intervals is chosen such that
reasonable statistics are guaranteed for each subset. An energy spectrum
is derived (see Section~\ref{subsec:DataVHE}) for each subset and is
subsequently fit by a power-law function with exponential cutoff
$\rm{d}N/\rm{d}E = I_{0} \cdot (E/1\,\rm{TeV})^{-\Gamma} \cdot
\exp(-E/E_{\rm{cut}})$. A fit of a simple power-law function can be
excluded with high confidence for most of the spectra (see as an example
the dotted line in Fig.~\ref{fig:VER_SpecFluxLevels}, resulting in a
$\chi^{2}/\rm{dof} = 264.3/15$). The results of the fits are summarized
in Table~\ref{tab:SpectrumTeV} and the energy spectra are shown in
Figure~\ref{fig:VER_SpecFluxLevels}. No correlation between the cutoff
energy $E_{\rm{cut}}$ and the flux normalization $I_{0}$ can be claimed. 
The spectra are also fitted with the same function by fixing the cut-off
energy to $E_{\rm{cut}} = 4 \, \rm{TeV}$ (Table~\ref{tab:SpectrumTeV}).
In this case, a hardening of the spectrum with increasing flux level can
be seen, see Figure~\ref{fig:VHE_FluxVsIndex}. A linear correlation
between the flux and the index $\Gamma$ is disfavored
($\chi^{2}/\rm{dof} = 44.6/5$, $p = 1.7 \cdot 10^{-08}$) as compared to
a quadratic relationship $\Gamma(I_{0}) = a + b \cdot I_{0} + c \cdot
I_{0}^{2}$ with $a = 2.70 \pm 0.04$, $b = (-5.3 \pm 0.6) \cdot 10^{9}$
and $c = (1.08 \pm 0.20) \cdot 10^{19}$ ($\chi^{2}/\rm{dof} = 13.2/4$,
$p = 0.01$). This finding indicates that the spectral hardening with
flux level flattens at very high flux values, as was seen already in the
case of PKS\,2155-304 \citep{PKS2155}. Given the sparse sampling during
most of the nights, we were not able to further separate the data into
rising and falling (with time) flux states which may have an effect on
the $\Gamma(I_{0})$ function. However, the general flux versus $\Gamma$
trend is in good agreement with earlier results obtained with the
Whipple $10\, \rm{m}$ telescope \citep{Krennrich2002} which are also
shown in Figure~\ref{fig:VHE_FluxVsIndex}.

\begin{table*}
\begin{center}

\caption{\label{tab:SpectrumTeV} Parameters and statistical errors of
fits $\rm{d}N/\rm{d}E = I_{0} \cdot (E/1\,\rm{TeV})^{-\Gamma} \cdot
\exp(-E/E_{\rm{cut}})$ to the energy spectra shown in
Figure~\ref{fig:VER_SpecFluxLevels}: observation time $T_{\rm{live}}$,
statistical significance of the excess in the data set, flux
normalization $I_{0}$, photon index $\Gamma$, cut-off energy
$E_{\rm{cut}}$ and the quality of the fit. The second row for each flux
state corresponds to a fit in which the cut-off energy was fixed to a
value of $E_{\rm{cut}} = 4 \, \rm{TeV}$.}

\begin{tabular}{lrrrrrr}
\tableline\tableline
 & $T_{\rm{live}} [\rm{h}]$
 & $\rm{sign.} [\sigma]$
 & $I_{0} \, [\frac{10^{-11}}{\rm{cm}^{-2} \, \rm{s}^{-1} \, \rm{TeV}^{-1}}]$
 & $\Gamma$
 & $E_{\rm{cut}} \, [\rm{TeV}]$
 & $\chi^{2}/\rm{dof}$ \\

\noalign{\smallskip}

\tableline

\noalign{\smallskip}

very-low-state
        & $ 7.83 $
	& $ 77.62 $
        & $ 4.78 \pm 0.73 $
        & $ 2.29 \pm 0.11 $
        & $ 1.59 \pm 0.34 $
        & $ 17.8 / 11 \, (1.62) $ \\

        &
	&
        & $ 3.056 \pm 0.086 $
        & $ 2.608 \pm 0.033 $
        & $ 4 \pm 0 $
        & $ 29.2 / 12 \, (2.44) $ \\

\noalign{\smallskip}

low-state
        & $ 16.3 $
	& $ 181.3 $
        & $ 7.60 \pm 0.33 $
        & $ 2.285 \pm 0.035 $
        & $ 2.95 \pm 0.29 $
        & $ 20.7 / 14 \, (1.48) $ \\

        &
	&
        & $ 6.769 \pm 0.084 $
        & $ 2.375 \pm 0.015 $
        & $ 4 \pm 0 $
        & $ 29.3 / 15 \, (1.95) $ \\

\noalign{\smallskip}

mid-state
        & $ 7.79 $
	& $ 160.6 $
        & $ 10.26 \pm 0.44 $
        & $ 2.278 \pm 0.037 $
        & $ 4.36 \pm 0.58 $
        & $ 14.9 / 15 \, (0.99) $ \\

        &
	&
        & $ 10.54 \pm 0.15 $
        & $ 2.256 \pm 0.017 $
        & $ 4 \pm 0 $
        & $ 15.4 / 16 \, (0.96) $ \\

\noalign{\smallskip}

high-state A
        & $ 1.38 $
	& $ 73.0 $
        & $ 19.08 \pm 2.63 $
        & $ 2.01 \pm 0.12 $
        & $ 1.91 \pm 0.41 $
        & $ 15.6 / 12 \, (1.30) $ \\

        & 
	&
        & $ 13.53 \pm 0.38 $
        & $ 2.295 \pm 0.037 $
        & $ 4 \pm 0 $
        & $ 24.0 / 13 \, (1.84) $ \\

\noalign{\smallskip}

high-state B
        & $ 0.63 $
	& $ 47.3 $
        & $ 22.23 \pm 2.42 $
        & $ 1.88 \pm 0.24 $
        & $ 3.06 \pm 1.07 $
        & $ 13.4 / 6 \, (2.24) $ \\

        &
	&
        & $ 20.77 \pm 1.14 $
        & $ 2.053 \pm 0.080 $
        & $ 4 \pm 0 $
        & $ 143.0 / 7 \, (2.00) $ \\

\noalign{\smallskip}

high-state C
        & $ 0.63 $
	& $ 53.76 $
        & $ 21.74 \pm 1.97 $
        & $ 2.40 \pm 0.26 $
        & $ 9.6 \pm 9.0 $
        & $ 0.9 / 6 \, (0.15) $ \\

        &
	&
        & $ 23.96 \pm 1.62 $
        & $ 2.05 \pm 0.09 $
        & $ 4 \pm 0 $
        & $ 2.7 / 7 \, (0.39) $ \\

\noalign{\smallskip}

very-high-state
        & $ 0.63 $
	& $ 63.5 $
        & $ 35.77 \pm 3.08 $
        & $ 1.87 \pm 0.17 $
        & $ 2.74 \pm 0.60 $
        & $ 6.2 / 9 \, (0.69) $ \\

        & 
	&
        & $ 32.0 \pm 1.2 $
        & $ 2.111 \pm 0.057 $
        & $ 4 \pm 0 $
        & $ 8.7 / 10 \, (0.87) $ \\


\tableline
\end{tabular}

\end{center}
\end{table*}

Mrk\,421 is detected by VERITAS (on average) with a statistical
significance well above 10~standard deviations per run, even for runs
with a duration of only 10~min. This enables a run-by-run derivation of
the energy spectrum on time intervals of 10~min or less. Any energy
spectrum derived from an individual data run which meets the following
requirements is fit by a power-law with exponential cutoff (the cutoff
energy again being fixed to $E_{\rm{cut}} = 4 \, \rm{TeV}$): (i) A
differential flux point is only considered in a fit if the statistical
significance of the excess is above 2~standard deviations; (ii) An
energy spectrum is only fit if at least four differential flux points
fulfill the first criterion. The results of the energy spectra derived
for the individual runs are shown in Figure~\ref{fig:VHE_FluxVsIndex}
(gray points) and confirm the trend which has been found already in the
data sets divided according to the different flux levels. 

For both cases, the correlation coefficient was calculated and the
corresponding non-directional chance probability for the null hypothesis
$p$ (non-correlation) was calculated using the Student
t-distribution\footnote{Note: The correlation factor does not account
for the statistical errors on the individual data points.}. The
correlation coefficients are $r = -0.86$ for the data sets separated by
flux level\footnote{Since this sample consists of only 7 data pairs no
chance probability $p$ was calculated for $r$.}, and $r = -0.55$ ($p <
10^{-4}$) for the distribution based on the run-by-run data sets.

\subsection{Temporal and Spectral Variability in the X-ray Band 
\label{subsec:Results_XRay}}

\paragraph{Energy spectra and fluxes.}

The RXTE/PCA spectra were fitted in an energy range of $3 - 20 \,
\rm{keV}$, while the Swift/XRT spectra were fitted between $0.4 - 10 \,
\rm{keV}$. Two models were tested to fit the data: a power law $\Phi(E)
= k E^{-\Gamma}$ and a log-parabolic model $\Phi(E) = k E^{-(a + b \log
(E/E_{0}))}$ \citep{Massaro2004, Massaro2006}. The log-parabolic
function uses an energy dependent photon index $\Gamma(E) = a + b \log
(E/E_{0})$; the parameter $b$ defines the curvature in the logarithmic
parabola, and $a$ is the spectral index at $E_{0}$. Both models account
for absorption assuming a fixed galactic column density of $1.61 \times
10^{20} \, \rm{cm}^{-2}$ \citep{LockSavage1995}. The mean reduced
$\chi^{2}$ values from log-parabolic fits to Swift/XRT and RXTE/PCA data
are $1.20$ ($96$~dof) and $0.87$ ($161$~dof), respectively. These are
significantly lower than the respective mean reduced $\chi^{2}$ values
from power law fits of 1.80 and 1.35. Hence, the integral flux from $2 -
10 \, \rm{keV}$ was calculated for each observation from log parabolic
fits.

\paragraph{Flux variability and light curves.}

The X-ray light curves of the $2 - 10 \, \rm{keV}$ fluxes are shown in
Figures~\ref{fig:MWL_LightCurveAll} to \ref{fig:MWL_LightCurveZoom3}.
Significant day time-scale flaring is seen in many months in both 2006
and 2008, while for the observations in 2007 the Mrk\,421 X-ray flux
remained relatively low. Of particular interest are the observations in
March to May 2008. RXTE/PCA recorded one of the highest ever X-ray
fluxes for Mrk\,421 on March 30, 2008 (phase~2a). However, during the
very high TeV $\gamma$-ray flux state measured by VERITAS in early May
2008 (phase~2b) the near-simultaneous X-ray flux was only found to be at
a moderate level.

For all but one case the RXTE observation exposures were shorter than 
1.5~hours (with an average exposure of $25 \, \rm{min}$). The Swift 
observations had a far larger range in exposures, from 15 minutes in 
many observations, to over 40 hours in each of seven observations; the 
average exposure per pointing was $3.3 \, \rm{hrs}$. Significant 
variability in both RXTE/PCA and Swift/XRT observations was found. The 
corresponding Swift observations spanned up to 3 days, with clear 
flaring in the rates on hour time-scales. Some of the RXTE/PCA light 
curves show a steady change in the count rate over 20 minute periods. 
However, more detailed studies of the sub-day flux variations in the 
X-ray data are beyond the scope of this paper.

\begin{figure*}[t]

\includegraphics[width=0.99\textwidth]{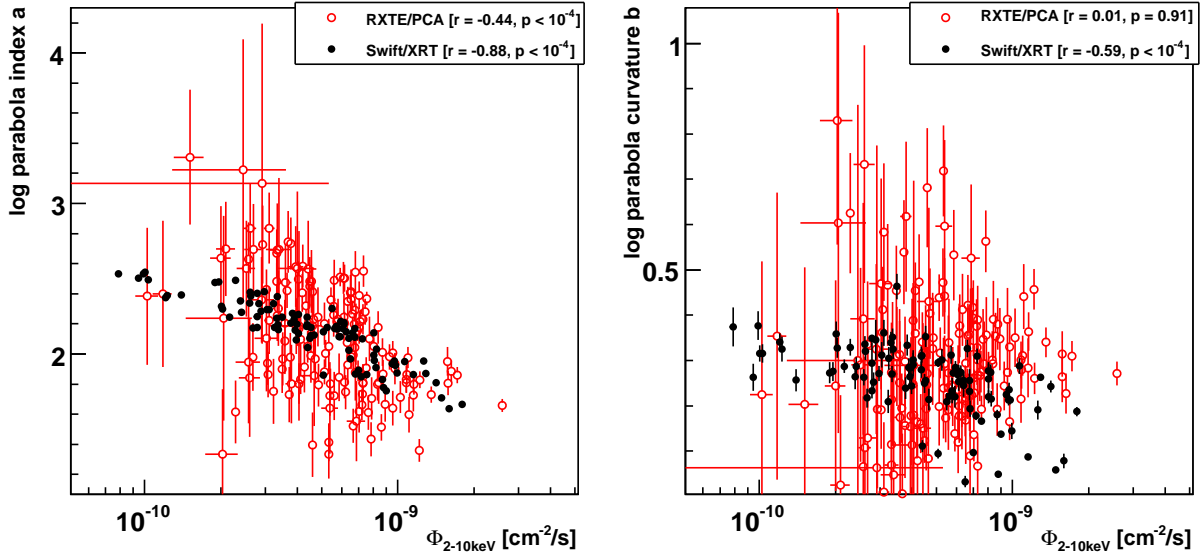}

\caption{\label{fig:XRay_FluxVsIndex} Correlations between the spectral
shape and the flux level for the X-ray data sets taken with RXTE/PCA and
Swift/XRT. Correlation coefficients $r$ and the corresponding chance
probability for the null hypothesis (no correlation) are given in the
legends. {\bf Left:} The log-parabola parameter $a$ vs. the $2-10 \,
\rm{keV}$ photon flux. {\bf Right:} The log-parabola curvature parameter
$b$ vs. the $2-10 \, \rm{keV}$ photon flux.}

\end{figure*}

\paragraph{Spectral variability.}

Drawing from this large set of X-ray observations, the correlation of
X-ray flux to spectral shape is investigated here. This will allow
identification of trends which could give important input to the
modeling. The left panel of Figure~\ref{fig:XRay_FluxVsIndex} shows the
correlation plot between the log-parabola parameters $a$ (index) vs. the
$2 - 10 \, \rm{keV}$ flux. The corresponding correlation plot between
the curvature parameter $b$ and the flux is shown in the right panel of
Figure~\ref{fig:XRay_FluxVsIndex}. Although the index $a$ is not
independent of the curvature parameter $b$, the conclusion can be
reached that an increased flux is accompanied by a hardening of the
spectrum (parameter $a$) since there is only a moderate correlation
between the flux and $b$. This finding is compatible with earlier
findings \citep{Fidelis2008}. The (anti)correlation between $a$ and the
flux is significant: the non-directional chance probability of the
derived correlation coefficient for the null-hypothesis is $p < 10^{-4}$
for the RXTE/PCA data, as well as for the Swift/XRT data.



\subsection{Flux Correlation Analysis \label{subsec:ResultsCross}}

This section describes the search for flux correlations between the
light curves measured in different energy bands and is based on the
light curves shown in Figures~\ref{fig:MWL_LightCurveAll} to
\ref{fig:MWL_LightCurveZoom3}.

\paragraph{Radio/optical/TeV flux correlations.}

\begin{table}
\begin{center}

\caption{\label{tab:OptFlares} Parameters of an exponential rise/fall
function (Sec.~\ref{subsec:ResultsCross}) $\Phi_{\rm{opt}}(t) = a + b /
(e^{-(t-T_{0}) / \tau_{\rm{r}}} + e^{(t-T_{0}) / \tau_{\rm{f}}})$ fitted
to the optical flares 'Opt\,1' and 'Opt\,2'
(Fig.~\ref{fig:MWL_LightCurveZoom}).}

\begin{tabular}{lrr}

\noalign{\smallskip}
\tableline\tableline
 Parameter & Opt\,1 & Opt\,2 \\

\noalign{\smallskip}
\tableline\tableline

\noalign{\smallskip}

$a$ [Jy]    & $0.021 \pm 0.001$ & $0.023 \pm 0.001$ \\
$b$ [Jy] & $0.023 \pm 0.002$ & $0.024 \pm 0.003$ \\
$\tau_{\rm{r}}$ [d] & $7.16 \pm 1.78$ & $9.52 \pm 1.35$ \\
$\tau_{\rm{f}}$ [d] & $5.46 \pm 0.80$ & $4.46 \pm 0.77$ \\
$T_{0}$ [MJD] & $54476.8 \pm 1.4$ & $54501.4 \pm 0.7$ \\

\tableline
\end{tabular}

\end{center}
\end{table}

As can be seen in Figure~\ref{fig:MWL_LightCurveAll} there is no
significant variation in the radio flux on day, week, or month time
scales. Thus, no significant correlation with fluxes measured in the
other wave bands is found. This could be explained if (i) the dominant
portion of the radio emission does not originate from the inner jet
region that likely produces the flux variability in the other energy
bands, or (ii) structures in the light curve are smeared out due to
slower cooling of the radio emitting electrons compared to the electrons
responsible for the shorter wavelength emission (see
Sec.~\ref{sec:discussion}).

\begin{figure}[t]

\epsscale{0.99}
\plotone{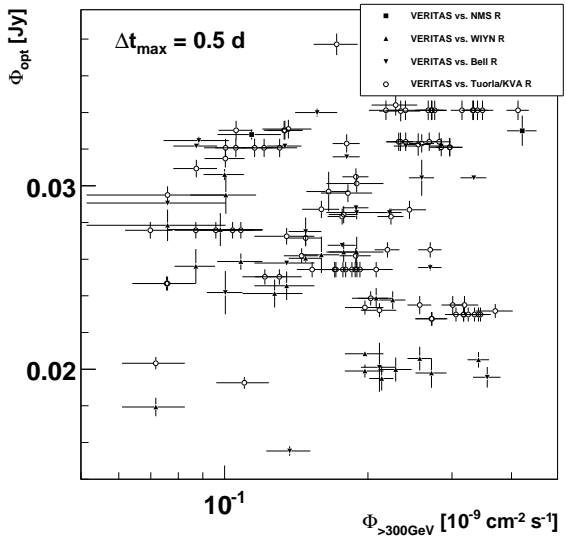}

\caption{\label{fig:Corr_VHE_Opt} Correlation between TeV and optical
fluxes. The maximum allowed time difference between the data points is
$\Delta t \leq 0.5 \, \rm{days}$.}

\end{figure}

The optical fluxes, on the other hand, show clear variability on time 
scales of weeks (Figure~\ref{fig:MWL_LightCurveZoom}), where the fluxes 
from the different bands (R, B, and V) are clearly correlated. This is 
not surprising since the frequency filter bands are not largely 
separated in terms of photon energy and a common origin of the radiation 
in the three bands can be assumed. Figure~\ref{fig:Corr_VHE_Opt} shows 
the correlation plot between the VERITAS TeV $\gamma$-ray fluxes and the 
optical fluxes in the R, B, and V bands. A maximum time gap between the 
TeV and optical measurements of $\Delta t \leq 0.5 \, \rm{days}$ was 
allowed for the individual data pairs. The corresponding correlation 
factors are compatible with chance expectations. A similar correlation 
study between X-ray and optical fluxes did not result in any significant 
correlation coefficient, either. Given the different time scales of the 
flux variations (sub-day level in the X-ray and TeV band, and weeks in 
the optical band), the lack of a direct flux correlation is expected.

Except for the Swift/UVOT data the contribution of the host galaxy is
not subtracted. Since it has to be constant in time, the light curves
can be used to set an upper limit on the base-line, which in turn would
be an upper limit on the host galaxy contribution in the measured data. 
This contribution is estimated to be $0.009 \, \rm{Jy}$ (R band, see
dotted line in Fig.~\ref{fig:MWL_LightCurveAll}), $0.009 \, \rm{Jy}$ (V
band), and $0.006 \, \rm{Jy}$ (B band). These estimates are somewhat
lower as compared the modeling of \citet{Nilsson1999} who estimate
$0.014 \, \rm{Jy}$ for the R band.

Although neither instantaneous nor delayed correlation between the
optical and the X-ray/TeV bands was found it is important to study the
structure of the optical light curves. Two well defined
flares\footnote{A third flare occuring around MJD $54149$ was also
fitted, resulting in comparable structural properties as listed in
Tab.~\ref{tab:OptFlares}.} 'Opt\,1' and 'Opt\,2'
(Figure~\ref{fig:MWL_LightCurveZoom}) were fitted with an exponential
rise/fall function $\Phi_{\rm{opt}}(t) = a + b /
(\exp(-\frac{t-T_{0}}{\tau_{\rm{r}}}) +
\exp(\frac{t-T_{0}}{\tau_{\rm{f}}}))$. The fit parameters are summarized
in Tab.~\ref{tab:OptFlares}, showing that the optical flux changes on
time scales of less than ten days with the indication of slightly
shorter fall times $\tau_{\rm{f}}$ as compared to the rise times
$\tau_{\rm{r}}$.

These time scales are on the same order as the estimated value of 
$\sim$$15 \, \rm{d}$, based on the characteristic X-ray/TeV variability 
time scale of $\leq 0.5 \, \rm{d}$ (see the discussion in 
Sec.~\ref{sec:discussion}). One can interpret $\Phi_{\rm{opt}}(t) $ as a 
'characteristic' optical response to a single X-ray/TeV flare. With this 
assumption a hypothetical prediction can be made for the optical light 
curve by folding $\Phi_{\rm{opt}}(t)$ (using $\tau_{\rm{r}} = 7 \, 
\rm{d}$ and $\tau_{\rm{f}} = 5 \, \rm{d}$, and a delay $\Delta T = 7 \, 
\rm{d}$) with the X-ray and/or TeV $\gamma$-ray light curves. The 
prediction based on the frequently sampled RXTE/ASM X-ray light curve 
(dwell-by-dwell, MJD $54450$ to $54630$, linear interpolation between 
the flux points) is shown together with the optical light curve in 
Figure~\ref{fig:MWL_LightCurveZoom} (arbitrary units)~-- no agreement is 
found. A caveat should be mentioned: the RXTE/ASM flux measurements are 
not very accurate; however, the general trends (high- vs. low-state) 
should allow the comparison of the two different wave bands. 
$\Phi_{\rm{opt}}(t)$ was also folded with the TeV $\gamma$-ray light 
curve (MJD $54500$ to $54630$, linear interpolation between the flux 
points). Again, no correlation can be found. There is a caveat here, as 
well: the non-continuous sampling of the TeV $\gamma$-ray light curve 
and the short duty-cycle of flares results in a considerable chance that 
one or more strong $\gamma$-ray flares have been missed which would 
change the shape of the folded optical light curve. Therefore, the 
latter results are not shown in Figure~\ref{fig:MWL_LightCurveZoom}. 
Furthermore, our above comparison ignores the possibility that particles 
with different energies are injected at the beginning of the flare.

\paragraph{Hard/soft X-ray flux correlations.}

\begin{figure}[t]

\epsscale{0.99}
\plotone{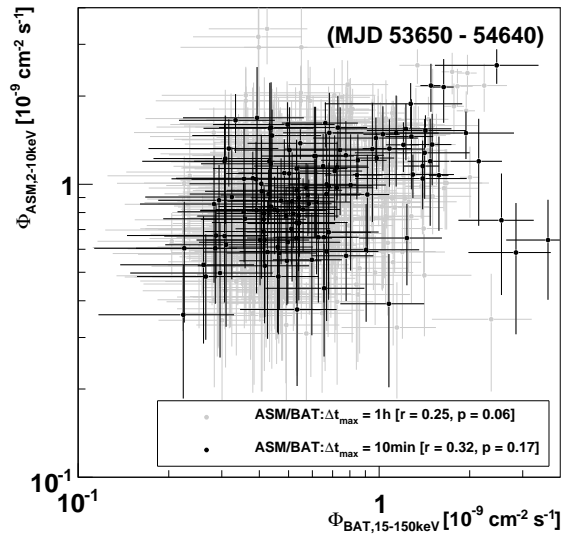}

\caption{\label{fig:XRay_BATvsASM} Hard/soft X-ray flux correlation
between the $15-150 \, \rm{keV}$ Swift/BAT flux and the $2-10 \,
\rm{keV}$ RXTE/ASM flux. All dwell-by-dwell flux points with a
statistical significance above $2 \, \sigma$ measured in the first half
of 2008 (MJD 54457--54640) are shown for two cases of a maximum time lag
between the observed data points. The legend gives the corresponding
correlation coefficient and the corresponding chance probability for the
non-correlation.}

\end{figure}

The correlation between hard and soft X-ray fluxes is studied based on
the dwell-by-dwell flux points measured with Swift/BAT ($15 - 150 \,
\rm{keV}$) and RXTE/ASM ($2 - 10 \, \rm{keV}$). The correlation plot
including all data points taken in spring 2008 (MJD 54457--54640,
contemporaneous with the VERITAS coverage) with a statistical
significance of more than $2$~standard deviations is shown in
Figure~\ref{fig:XRay_BATvsASM}. The distribution was generated for two
different requirements regarding the maximum allowed time gap between
the center time of the individual pointings ($\Delta t \leq 1 \, \rm{h}$
and $\Delta t \leq 10 \, \rm{min}$). In both cases, indications for only
a weak correlation of $r \simeq 0.1$ are found with moderate
significance. However, while the RXTE/ASM data are testing the falling
edge of the synchrotron peak in the SED, the Swift/BAT data likely fall
into the transition zone between synchrotron and high-energy peaks,
compare with Figures~\ref{fig:SEDs1} and \ref{fig:SEDs2}.

\paragraph{X-ray/TeV flux correlations.}

\begin{figure}[h]

\epsscale{0.99}
\plotone{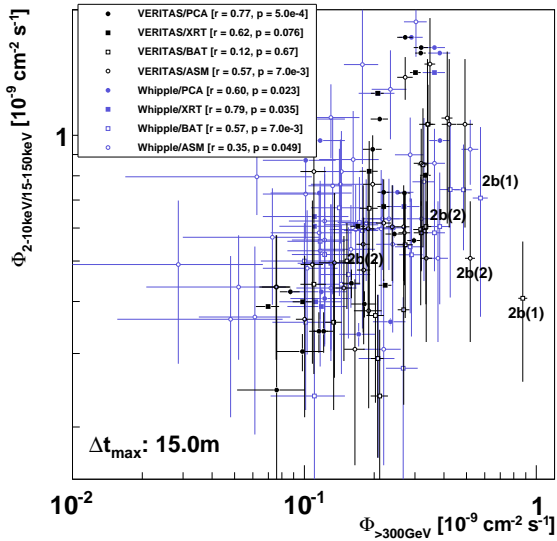}

\caption{\label{fig:Corr_VHE_Xray} Correlation between TeV fluxes
(VERITAS and Whipple) and X-ray fluxes (RXTE: PCA \& ASM, Swift: XRT \&
BAT). All X-ray fluxes are given between $2 - 10 \, \rm{keV}$, except
for the Swift/BAT fluxes which are given between $15 - 150 \, \rm{keV}$.
Only data measured in the first half of 2008 (MJD 54457--54640) are
shown. The RXTE/ASM and Swift/BAT fluxes are dwell-by-dwell with the
requirement that the statistical significance is above $2$ standard
deviations. The maximum time difference between the data points is
$\Delta t \leq 15 \, \rm{min}$. The correlation coefficients including
the chance probability for the null hypothesis are shown in the legend. 
The flux pairs measured during the first (2b(1)) and second night
(2b(2)) of phase~2b are marked.} 

\end{figure}

The rich data sets allow for a detailed study of the X-ray versus TeV
flux correlations. The TeV runs have a duration of $10-20 \, \rm{min}$.
With a few exceptions (during the highest flare) no indications for TeV
flux variations were found within any of the runs. A maximum time lag
between the correlated X-ray/TeV data points of $\Delta t \leq 15 \,
\rm{min}$ was allowed (close to the average TeV run duration). Since
some of the dedicated, high-quality data from Swift/XRT and RXTE/PCA
were considerably longer in exposure, an additional cut on the total
length of the X-ray pointing was applied (based on the average exposure
per pointing): $\Delta T \leq 30 \, \rm{min}$ in the case of RXTE/PCA
and $\Delta T \leq 2 \, \rm{h}$ in the case of Swift/XRT. Flux
variations within individual X-ray pointings are not investigated in the
framework of this paper; therefore, the data pairs cannot be considered
as exactly simultaneous. The RXTE/ASM and Swift/BAT dwell-by-dwell data
points have exposures of the order of $1-10 \, \rm{min}$, so that no cut
on $\Delta T$ was applied in those cases. Individual RXTE/ASM and
Swift/BAT flux measurements were only considered in case of a
significance level of $\geq 2 \, \sigma$. The results discussed below do
not strongly depend on the exact choice of $\Delta t$ and $\Delta T$.

Figure~\ref{fig:Corr_VHE_Xray} shows the flux correlations for the data
taken in spring 2008 (MJD 54457--54640). The correlation coefficients
and the corresponding non-directional chance probability for the null
hypothesis (no correlation) were calculated and are shown in the figure
legend. The X-ray and TeV fluxes seem to be correlated ($r > 0.5$ for
most data subsets) with chance probabilities of the order of a few
percent or below. The VERITAS/BAT data set is the only one which is not
correlated in a significant manner. One of the high TeV flux points
measured during the strong TeV $\gamma$-ray flare
(Figure~\ref{fig:MWL_LightCurveZoom3}) was accompanied by a Swift/BAT
dwell-by-dwell pointing ($\Delta t \leq 3.4 \, \rm{min}$), which however
did not indicate an increased activity in the hard X-ray band. The
Swift/BAT point has a duration of $8 \, \rm{min}$ and is therefore fully
contained in the time intervall of the corresponding VERITAS data run of
$20 \, \rm{min}$ duration. This flux pair is located in the lower right
corner of Figure~\ref{fig:Corr_VHE_Xray} (labeled as 2b(1)) and may be
seen as the indication of an orphan TeV $\gamma$-ray flare. However,
since this indication is based on only one X-ray/TeV flux pair measured
during a high TeV $\gamma$-ray flux state, there is not enough evidence
for a strong claim. If the corresponding data pair were removed, the
VERITAS/BAT flux correlation would increase to $r=0.62$ with a chance
probability of $p=0.019$. Other flux pairs from the two TeV flare nights
(phase~2b) are indicated in Figure~\ref{fig:Corr_VHE_Xray}, as well.

\paragraph{Discrete correlation functions.}

The light curves were also analyzed using the discrete correlation
function (DCF) technique \citep{Edel:88} in order to search for
correlated/delayed emission between different energy bands, allowing for
significant time lags (i.e. one light curve having the delayed shape of
another one). The radio, optical and X-ray light curves were tested
against the TeV $\gamma$-ray light curve. Except for the zero lag
X-ray/TeV correlation (see above) no significant time lag and/or
correlation at zero time lag was found for the radio and optical band as
compared to the TeV $\gamma$-ray emission.

\subsection{Spectral Energy Distributions and Modeling
\label{subsec:SEDs}}

\begin{figure*}[p]

\centering{
\includegraphics[width=0.99\textwidth]{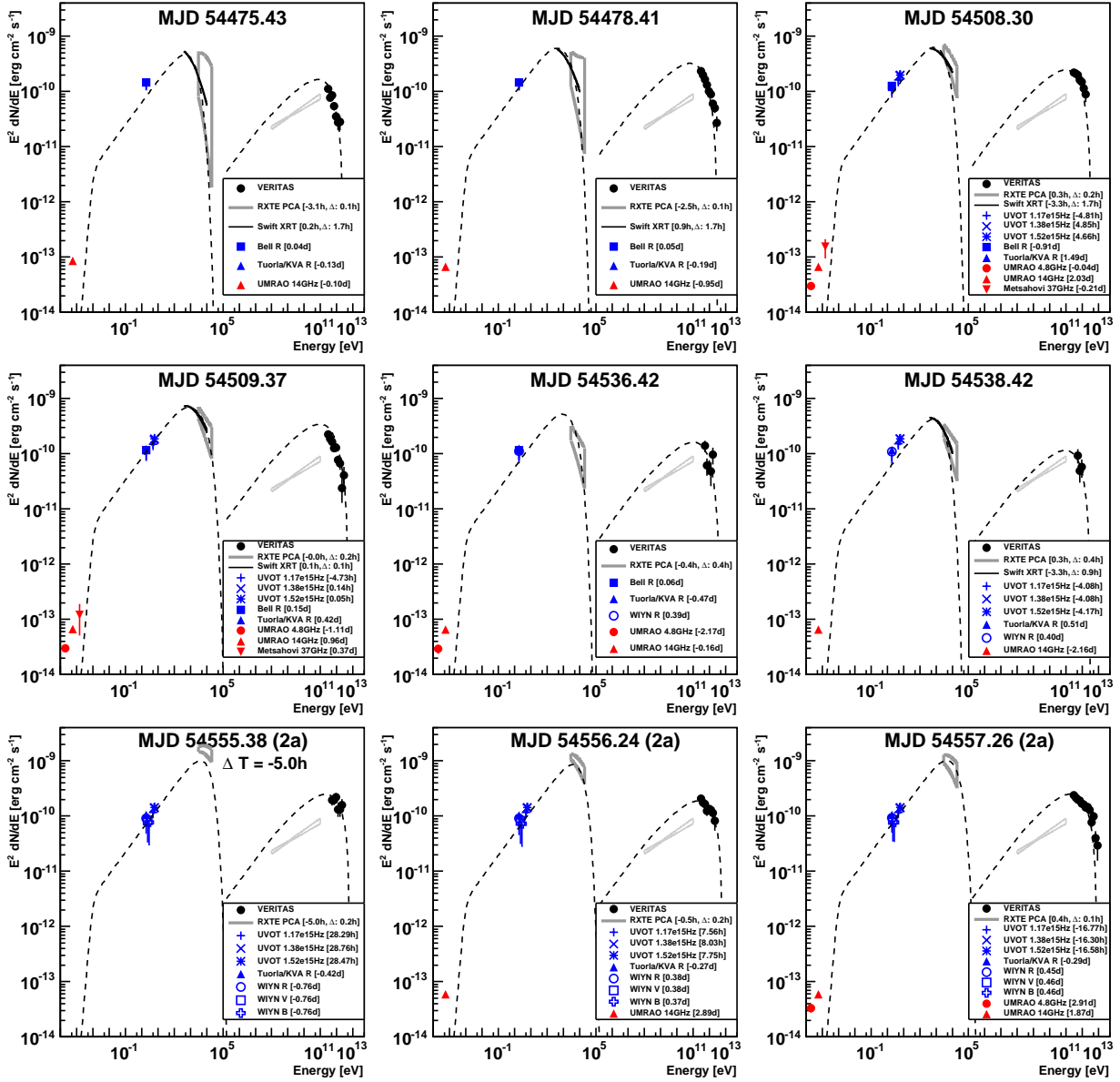}
}

\caption{\label{fig:SEDs1} SEDs of Mrk\,421 for individual dates showing
all available MWL data (statistical errors only) if the corresponding
time lag criterion is met (see text for more detail). The legends
indicate the time lags $\Delta T = T_{\rm{MWL}} - T_{\rm{TeV}}$ between
the individual MWL data and the VERITAS data. For the X-ray
observations, the duration of the pointing $\Delta$ is given, as well.
The number in parentheses refers to the phase (2a or 2b) during which
the The SEDs were fit by a one-zone SSC model (dashed curves); the
corresponding model parameters are summarized in
Table~\ref{tab:SSC_Params}. The non-simultaneous MeV/GeV spectrum
measured by Fermi (August 2008 to July 2009) is shown for reference
\citep{FirstFermiCatalog}.}

\end{figure*}

\begin{figure*}[p]

\centering{
\includegraphics[width=0.99\textwidth]{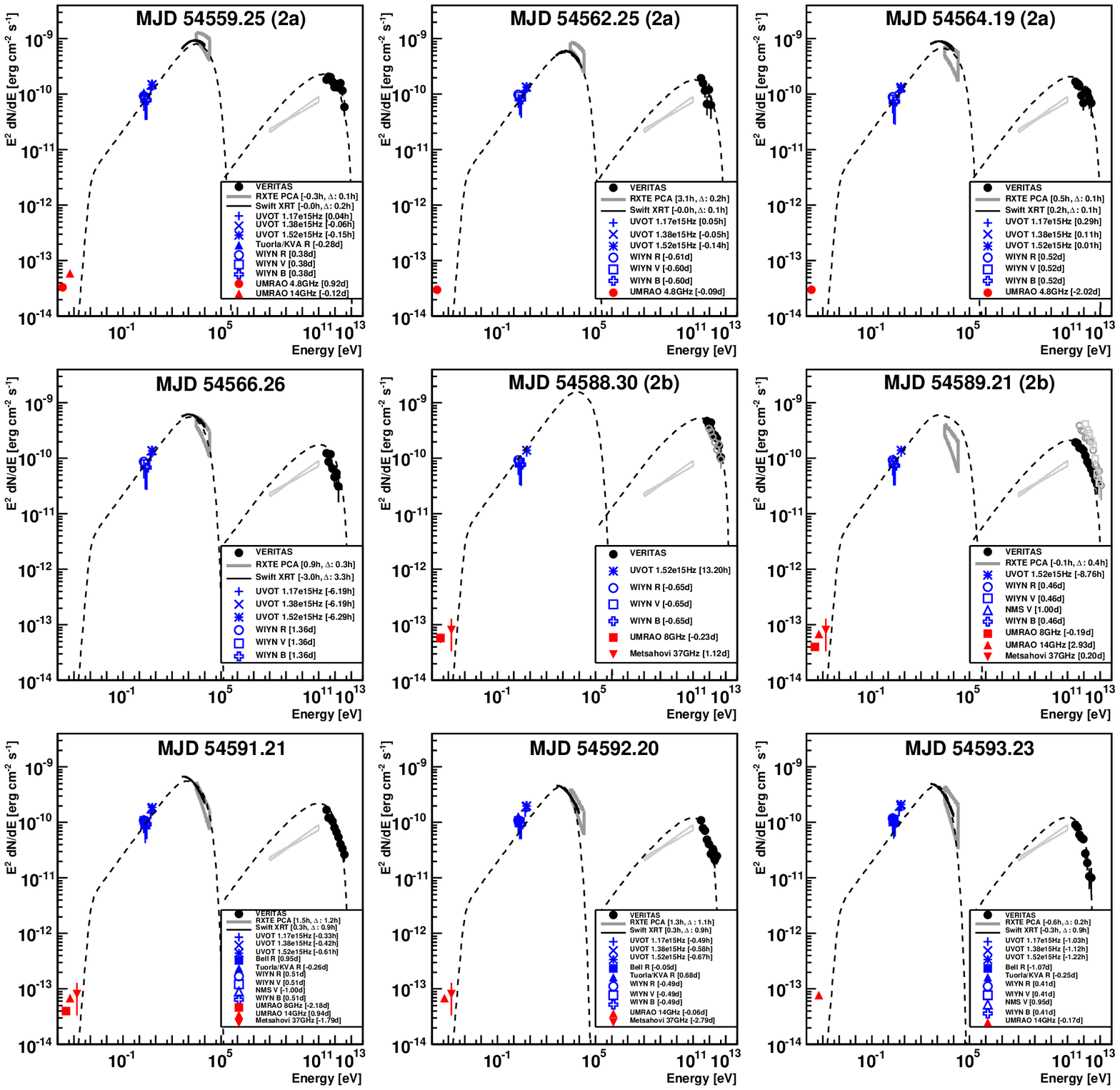}
}

\caption{\label{fig:SEDs2} SEDs of Mrk\,421, continued from 
figure~\ref{fig:SEDs1}. The SEDs of the two flare nights are also shown: 
MJD 54588.30 (no X-ray coverage in this night) and MJD 54589.21 with 
X-ray coverage during the onset of the flare (solid TeV points) but not 
during the high flare state (open TeV data points).}

\end{figure*}

The spectral energy distributions (SEDs) showing the VERITAS and
multi-wavelength (MWL) data were generated for individual nights and
plotted in Figures~\ref{fig:SEDs1} and \ref{fig:SEDs2}. Based on the
different variability time scales, data from the different wavebands are
plotted in the quasi-simultaneous SEDs if the time lag $\Delta t$
between the MWL data and the TeV data is $\Delta t_{\rm{X-ray}} \leq
0.15 \, \rm{d}$ (X-ray), $\Delta t_{\rm{opt}} \leq 1.5 \, \rm{d}$
(optical), and $\Delta t_{\rm{radio}} \leq 3 \, \rm{d}$ (radio). Shorter
time lags are present (down to real simultaneity) and the exact times
are given in the figure legends of the SEDs. The time spans (duration)
of the X-ray observations are given in the legends, as well. Possible
spectral variations within individual X-ray observations are beyond the
scope of this paper and are ignored in the modeling of the SEDs. The
X-ray spectra (Swift/XRT and RXTE/PCA) were fit with a log-parabola
model $\Phi(E) = k E^{-(a+b \log (E))}$, and the corresponding
log-parabola bow ties including the statistical errors of the fits are
shown if the fit resulted in a $\chi^{2}/\rm{dof} < 2$.  All errors are
statistical only.

As a reference, Figures~\ref{fig:SEDs1} and \ref{fig:SEDs2} also show
the non-simultaneous MeV/GeV energy spectrum measured by the Fermi/LAT
\citep{FermiLAT} in August 2008, to July 2009 \citep{FirstFermiCatalog}.
During this period, the LAT detects Mrk\,421 with a statistical
significance of $88 \, \sigma$ and the energy spectrum is well fit by
a power law in the whole energy range (curvature index given in the
Fermi catalog of $0.72$). Interestingly, the variability index of $43.9$
does not indicate very strong variability during the above period, so
that the Fermi/LAT energy spectrum shown in the SEDs may give a
realistic indication for a low/medium flux state of Mrk\,421. However,
given the non-simultaneity of the Fermi observations, the MeV/GeV
spectrum was not included in the modeling.

The SEDs were fit with a one-zone SSC model following
\citet{Krawczynski_1ES1959}. In this model a spherical emission region
of radius $R$ is filled with a relativistic electron population,
traveling down the jet with a bulk Lorentz factor $\Gamma$ ($\beta =
v/c$). The magnetic field $B$ in the emission region is randomly
oriented. The emitted radiation is Doppler-shifted by $\delta =
[\Gamma(1 - \beta \cos \theta)]^{-1}$, $\theta$ being the angle between
the jet axis and the line of sight of the observer. The electron
distribution is normalized by a factor $u_{\rm{e}}$ (in units of
ergs/cm$^{3}$) and is described in the jet frame by a broken power law
with $E_{\rm{min}}$, $E_{\rm{brk}}$, and $E_{\rm{max}}$ and the two
corresponding spectral indexes $\alpha_{1} = -2.2$ and $\alpha_{2} =
-3.2$. Mrk\,421 is located at a redshift of $0.031$ so that the
energy-dependent pair absorption on photons of the extragalactic
background light (EBL) cannot be neglected.  EBL absorption is taken
into account in the model following Franceschini et al. (2008). Since
most of the optical data were not corrected for the contribution of the
host galaxy, the optical data points have to be seen as an upper limit
on the jet emission region. The estimate on the host galaxy contribution
(Sec.~\ref{subsec:ResultsCross}) has been added as a systematic error.
The SSC model was adjusted to the SEDs with the following procedure:

\begin{itemize}

\item In a first step, the three SEDs with the smallest time lags
between the X-ray and the TeV spectra (MJD 54559.25, 54562.25, and
54509.37) were used to derive a full set of model parameters. The
Doppler factor was set to $\delta = 40$ and the magnetic field was set
to $B = 0.2 \, \rm{G}$, in agreement (order of magnitude) with a
synchrotron cooling time of $\tau_{\rm {sync}}(B) \times \delta \,
\approx$ $T_{\rm{flare}}\, = 1 \, \rm{h}$ for the electrons emitting the
synchrotron emission close to the maximum of the spectral energy
distribution. Then, the radius $R$, the energy density in electrons
$u_{\rm{e}}$ [ergs/cm$^{3}$], as well as $E_{\rm{brk}}$ and
$E_{\rm{max}}$ were varied until the model would describe the SED
(optical to TeV).

\item In a second step, the remaining SEDs were fit by only allowing 
variations of $u_{\rm{e}}$ and $E_{\rm{max}}$ (and in case of bad fits 
also $E_{\rm{brk}}$) as compared to the model parameters derived above, 
reflecting a change in the injected electron population.

\end{itemize}

The SSC model fits are shown in Figure~\ref{fig:SEDs1} and
\ref{fig:SEDs2} and the model parameters for the individual SEDs are
summarized in Table~\ref{tab:SSC_Params}. The models generally
under-predict the radio emission which is synchrotron self-absorbed in
the model at the low-energy tail of the SED. This discrepancy could be
explained by additional radio emission from regions in the jet not
emitting the TeV radiation. Also, the synchrotron self-absorbed radio
blobs could expand and lead to a delayed radio emission from a larger
region \citep{M87Joint}, which however is not taken into account in the
model.

The break energy $E_{\rm{brk}}$ of the electron spectrum is very near 
the maximum energy of the electron spectrum $E_{\rm{max}}$. The missing 
plateau ($E_{\rm{brk}} \sim E_{\rm{max}}$) is directly seen in the 
Swift/XRT data which show a direct turn-over of the synchrotron peak. 
This is in agreement with earlier findings in the case of Mrk\,421 
\citep{Fossati2008} and is in contrast to the SEDs measured in the case 
of Mrk\,501 for which the synchrotron emission peaks at higher energies 
and shows indications of a plateau \citep{Krawczynski2000}. If 
$E_{\rm{min}} \ll E_{\rm{brk}}$, the missing plateau implies that 
cooling did not have time to kick in. However, if $E_{\rm{min}} = 
E_{\rm{brk}}$ and cooling is very efficient, electrons cool below 
$E_{\rm{brk}}$ and one gets a new 'effective' $E_{\rm{min}} \ll 
E_{\rm{brk}}$, and the 'true' $E_{\rm{min}}$ becomes $E_{\rm{brk}}$.

All SEDs are reasonably described by using the same model parameters 
except for the electron normalization $u_{\rm{e}}$ and the break/maximum 
electron energies $E_{\rm{brk}}$ and $E_{\rm{max}}$ which were adjusted 
for each individual SED (Table~\ref{tab:SSC_Params}). The only 
exceptions are the X-ray flare (Figure~\ref{fig:SEDs1}) and the two TeV 
$\gamma$-ray flare SEDs (Figure~\ref{fig:SEDs2}):

\begin{itemize}

\item MJD 54555.38: The RXTE/PCA bowtie shown in this SED
(Figure~\ref{fig:SEDs1}) does not qualify for the previously defined
X-ray/TeV time lag criterion. However, the X-ray flux is the highest one
measured during the whole MWL campaign (see dotted line in
Figure~\ref{fig:MWL_LightCurveZoom2}) so that it is shown for reference
(but not included in the fit). The closest Whipple TeV flux point is
$\sim$$2 \, \rm{h}$ away and also does not show signs for an increased
TeV flux. This may indicate an orphan X-ray flare, but the lack of
simultaneous TeV data does not allow a strong conclusion.

\item MJD 54588.3: The SSC model slightly over-predicts the optical
emission. However, the optical data are not simultaneous and the TeV
fluxes can change within $20 \, \rm{min}$: the black data points in
Figure~\ref{fig:SEDs2} represent the first $20 \, \rm{min}$ of the
flare, whereas the open gray points represent the second $20 \,
\rm{min}$ of the flare in that night (compare with the lower left insert
in Figure~\ref{fig:MWL_LightCurveZoom3}). Unfortunately, no X-ray data
(Swift/XRT or RXTE/PCA) were taken during this night.

\item MJD 54589.21: The black data points in Figure~\ref{fig:SEDs2}
represent the low plateau at the beginning of the flare~-- compare with
the lower right insert in Figure~\ref{fig:MWL_LightCurveZoom3}~-- and
are the ones which were used for the fit of the SED. The RXTE/PCA
measurement partially overlapped this time, but is over-predicted by the
model. Given the generally good fits of the model, this is interesting
since the VERITAS/BAT flux pairs of the same night (see 2b(2) in
Figure~\ref{fig:Corr_VHE_Xray}) as well as the previous night (2b(1))
also indicate a possible orphan TeV $\gamma$-ray flare. The higher flux
states during this flare night (open gray points in
Figure~\ref{fig:SEDs2}) were not accompanied by simultaneous X-ray
measurements and were therefore not modeled.

\end{itemize}

The fact that the X-ray (and optical) data are not perfectly described 
during the flare nights (as well as for a few other nights) may be 
explained by the fact that only $u_{\rm{e}}$, $E_{\rm{brk}}$ and 
$E_{\rm{max}}$ were allowed to vary, after the other parameters had been 
fixed based on the three most complete and contemporaneous SEDs, which 
all correspond to TeV low/medium states of Mrk\,421.

\begin{table}[t]
\begin{center}

\caption{\label{tab:SSC_Params} Parameters of the SSC models as shown in 
Figure~\ref{fig:SEDs1} and \ref{fig:SEDs2}. The following parameters are 
the same for all models: Doppler factor $\delta = 40$, magnetic field $B 
= 2.0 \times 10^{-5} \, \rm{T} = 0.2 \, \rm{G}$, radius $R = 2.5 \times 
10^{15} \, \rm{cm}$, $\log (E_{\rm{min}}) = 3.0$. The table shows the 
remaining parameters $u_{\rm{e}}$, $\log (E_{\rm{max}})$ and $\log 
(E_{\rm{brk}})$ for the different SEDs, as well as the particle energy 
density to magnetic field energy density ratio $u_{\rm{e}} / u_{\rm{B}} 
= 8 \pi u_{\rm{e}} / B^{2}$.}

\begin{tabular}{lrrrr}

\noalign{\smallskip}
\tableline\tableline
 date &
	$u_{\rm{e}}$ &
	$\log (E_{\rm{max}})$ &
	$\log (E_{\rm{brk}})$ &
	$u_{\rm{e}} / u_{\rm{B}}$ \\

 [MJD] & [ergs/cm$^{3}$] & & & \\

\noalign{\smallskip}
\tableline\tableline

\noalign{\smallskip}

54475.4 & 0.45 & 10.8 & 10.5 & 283 \\
54478.4 & 0.65 & 10.9 & 10.4 & 408 \\
54508.3 & 0.50 & 11.0 & 10.6 & 314 \\
54509.3 & 0.60 & 11.1 & 10.5 & 377 \\
54536.4 & 0.40 & 11.0 & 10.6 & 251 \\
54538.4 & 0.35 & 10.9 & 10.6 & 220 \\
54555.4 & 0.40 & 11.2 & 11.2 & 251 \\
54556.3 & 0.35 & 11.2 & 11.1 & 220 \\
54557.3 & 0.40 & 11.3 & 11.0 & 251 \\
54559.2 & 0.40 & 11.4 & 10.8 & 251 \\
54562.2 & 0.40 & 11.3 & 10.6 & 251 \\
54564.2 & 0.40 & 11.4 & 10.7 & 251 \\
54566.2 & 0.40 & 11.2 & 10.6 & 251 \\
54588.3 & 0.55 & 11.6 & 11.0 & 345 \\
54589.3 & 0.43 & 11.5 & 10.6 & 270 \\
54591.3 & 0.48 & 11.2 & 10.5 & 301 \\
54592.3 & 0.35 & 11.0 & 10.6 & 220 \\
54593.3 & 0.35 & 11.0 & 10.6 & 220 \\

\tableline
\end{tabular}

\end{center}
\end{table}

\section{Summary and Discussion \label{sec:discussion}}

Together with many MWL partners VERITAS conducted an intensive MWL
campaign on Mrk\,421 in 2008. During low states of Mrk\,421, VERITAS is
able to measure the source flux to an accuracy of $5\%$ in $20 \,
\rm{min}$ time intervals. In the $47 \, \rm{hr}$ data set presented in
this paper, we did not find evidence for rapid flux variability on time
scales of minutes as reported for Mrk\,501 \citep{Mrk501MAGIC} and
PKS\,2155-304 \citep{PKS2155}. However, the only strong outburst
(reaching a flux level of 10~Crab) which would have allowed testing
these short time scales of flux variations was measured at high zenith
angles with a strongly reduced detection sensitivity.

Two phases of enhanced X-ray (phase~1) and X-ray/TeV (phase~2) activity
are found. Phase~2 can be separated into a period of strong X-ray
activity without a strong TeV $\gamma$-ray flaring, followed by a TeV
$\gamma$-ray flare lasting for two days without any indication for
contemporaneous strong X-ray activity. This may indicate an {X-ray
(phase~2a) and TeV $\gamma$-ray (phase~2b)} orphan flare, but the data
are too sparse for a definite claim. In the remaining data there is
significant evidence for a correlation between the X-ray and TeV fluxes.

No significant flux correlations between the TeV band and the
optical/radio bands were found. Assuming that (i) the optical and
X-ray/TeV photons are emitted co-spatially, and that (ii) the flux
variability time scale equals the radiative cooling time, one can
estimate the expected relation between the observed time scales of flux
variations: The mean energy of emitted photons $E_{\rm{obs}}$ scales
with the electrons' Lorentz factor squared, $E_{\rm{obs}} \propto
\gamma^{2}$. The cooling time $\tau_{\rm{cool}}$ scales proportional to
$1/\gamma$. The energy of the X-ray photons is three orders of magnitude
higher than the energy of the optical photons. The Lorentz factors of
the X-ray emitting electrons should thus be $1.5$ order of magnitudes
(factor $\sim$$30$) larger than the Lorentz factors of the optically
emitting electrons. If the X-rays (and $\gamma$-rays, from inverse
Compton scatterings of the electrons which emit X-rays as synchrotron
emission) show flux variability on a time scale of $\leq 0.5 \, \rm{d}$
\citep{Fidelis2008, Gaidos1996}, then the optical fluxes should vary on
a time scale of $\leq 15 \, \rm{d}$ which is well compatible with the
observed time scales in the optical and X-ray/TeV bands
(Sec.~\ref{subsec:ResultsCross}). However, as discussed in
Sec.~\ref{subsec:ResultsCross}, the optical light curve does not seem to
be a delayed and stretched version of the X-ray light curve or the TeV
$\gamma$-ray light curve, implying, that the dominant fraction of the
observed optical emission does not originate from the X-ray/TeV emission
region. Iterestingly, \citet{Rieger2004} discusses variable (even
periodic) flux variations on (periodic) time scales of $P \leq 10 \,
\rm{d}$ which can be explained by geometrical arguments of internal jet
rotation. Our data, however, seems to indicate more complicated
structures than periodicity. A similar estimate using the radiative
cooling time for the energies in the radio band leads to time scales of
$10$ years or more, impossible to test with the given MWL data set.

Clear indications for spectral hardening with increasing flux levels are
found in the X-ray and TeV bands. In the TeV band, the spectral
hardening seems to level out for the very high fluxes above
$\sim$5~Crab. A similar trend had already been found in earlier Whipple
data of Mrk\,421, as well as in the strong flare of PKS\,2155-304
measured by H.E.S.S. \citep{PKS2155}.

The rich MWL data set presented in this paper allowed for the 
compilation of $18$ quasi-simultaneous SEDs, well constrained by 
accurate Swift/XRT X-ray and VERITAS TeV spectra. The SEDs can be 
described by a one-zone SSC model with nearly one set of parameters; 
only $u_{\rm{e}}$, $E_{\rm{brk}}$ and $E_{\rm{max}}$ needed to be 
adjusted to describe the whole set of SEDs. However, for most of the 
SEDs Mrk\,421 was found to be in low or moderate flux states, leading to 
similar SEDs on different days. Our SSC modeling indicates that the 
emission process and the jet parameters are reasonably well constrained.

So far, there are only four TeV blazars with a reasonable amount of
simultaneous MWL data in order to claim a correlation between X-ray and
TeV $\gamma$-ray fluxes: Mrk\,421 \citep{Fossati2008, Blazejovski2005,
Horan2009}, Mrk\,501 \citep{Krawczynski2002}, PKS\,2155-304
\citep{PKS2155}, and 1ES\,2344+512 \citep{VER_2344}. Such a correlation
implies that the same high-energy particle population (e.g. electrons)
is responsible for the synchrotron emission at X-ray energies, as well
as the high-energy IC emission at TeV energies, as predicted in the
framework of SSC models. Investigating the exact shape of the X-ray/TeV
correlation (linear, quadratic, etc.) will be one of the important goals
for future studies.  Although this correlation is seen as a general
trend, it does not necessarily hold true at the level of individual
flares \citep{Krawczynski_1ES1959}. In our data, we find indications of
an X-ray high-state not accompanied by TeV $\gamma$-ray flaring
(phase~2a), as well as a TeV $\gamma$-ray flare without increased X-ray
activity (phase~2b). Although the data is not exactly contemporaneous~--
not allowing for a firm conclusion~-- such orphan flares in general
would require fine tuning of the SSC model \citep{Krawczynski_1ES1959}
or alternative models, e.g. external-Compton models or models where the
$\gamma$-ray emission is produced by hadrons, e.g. as proton-synchrotron
emission \citep{Muecke_SPB, Aharonian_Protons} or through a proton
induced cascade \citep{Mannheim_ProtonsMrk501}. Further observations are
needed to understand this particular aspect of the TeV flaring activity.

\citet{Acciari_Mrk421} reported comparable $\sim$week-scale trends
between the Mrk\,421 X-ray and optical fluxes without a strong X-ray/TeV
coupling. \citet{PKS2155_Opt} reported a clear indication of an
optical/TeV flux correlation in case of PKS\,2155-304. However, in the
second PKS\,2155-304 flare the optical/TeV correlation was not seen
although the data were again strictly simultaneous \citep{PKS2155}. 
Also, an optical/TeV correlation is not found in the large data sample
presented in this paper or in earlier large data sets. Therefore, it
does not seem to be a general property of TeV blazars. This may indicate
that (part of) the optical emission is dominated by a region larger than
the TeV emission site and/or a different emission mechanism is at play.
However, a certain level of optical synchrotron emission is un-avoidable
given the synchrotron emission at X-ray energies.  The ambiguous
findings so far in terms of the optical/TeV correlation may further
indicate that different emission scenarios may play a role in different
situations, e.g. we do not always observe the same type of flares. No
correlation between TeV and radio fluxes has been established at this
point resulting in similar arguments as above concerning the emission
regions/mechanisms.

TeV flux variations are measured for different TeV blazars with
characteristic time scales down to $5-20 \, \rm{min}$
\citep{Mrk501MAGIC, PKS2155, Gaidos1996}. Interestingly, the
corresponding size of the emission region reaches down to the order of
the Schwarzschild radius of the black hole of the corresponding AGN
\citep{Mrk501MAGIC, PKS2155}. This can be seen as an indication that the
TeV $\gamma$-ray emission from blazars comes from the base of the jet
where the jet energy density is highest and the jet cross section is
smallest. A similar finding was made in case of the radio galaxy M\,87
for which a promising approach to locate the site of the TeV emission
region was presented based on the combination of TeV $\gamma$-ray
observations with simultaneous high-resolution radio observations
\citep{M87Joint}.

We see two paths towards improving our understanding of the inner
workings of AGN jets. The first path involves simultaneous TeV
observations with imaging telescopes with $\leq 1 \, \rm{mas}$ angular
resolution, e.g. VLBA \citep{M87Joint}, and/or with polarimetric
observations in the optical \citep{Marscher08} and the X-ray
band\footnote{see for example {\it
http://heasarc.gsfc.nasa.gov/docs/gems/}.}. Furthermore, future
observations in the MeV/GeV band with Fermi, together with observations
in the GeV/TeV band with VERITAS, will facilitate the study of the
spectral slope, time scales and correlation of the fluxes in both energy
bands which are important inputs for the theoretical modeling. The
second path for achieving further progress concerns the theoretical
modeling of the results. The analysis presented in this paper confirms
that SSC models are successful in describing the broad-band spectral
energy distributions of high-frequency peaked BL Lac objects. The two
lessons which we infer from the modeling are (i) a large value of the
relativistic Doppler factor $\delta$ is required to explain the SEDs and
the rapid flux variability \citep{Gaidos1996,Bege:08}; (ii) in the case
of SSC models, the electron energy density exceeds the magnetic field
energy density both measured in the rest frame of the emitting volume
(see Table~\ref{tab:SSC_Params}). In the case of EC models, one can find
models with approximately equal electron and magnetic field energy
densities (e.g. \citet{Ghis:09}), or models in which this ratio deviates
considerably from unity \cite{Krawczynski2002}. Considering that protons
may add to the particle energy density, the particle energy density may
still be an equally comparable component in the plasma blobs producing
HBL flares. Further progress can be achieved by combining them with
those from theories describing the formation and structure of jets. 
Recently, general relativistic magneto-hydrodynamic simulation codes
(e.g.  \citet{McKi:06, Komi:07, Krol:10, Spru:10}) have been used to
validate aspects of analytic models of the magnetic formation,
acceleration, and collimation of jets \citep{Webe:67, Blan:77, Phin:83,
Came:86, Love:87, Li:92, Vlah:04, Kraw:07}. Giannios et al. 2009
discussed the model of "mini-jets in a jet" driven by magnetic
reconnection. Mini-jets in jets may be able to reconcile the predictions
of magnetic models of jet formation with the results from SSC (or EC)
modeling of the blazar emission of jets: The reconnection mechanism
converts magnetic energy into particle energy
\citep{Siko:05,Gian:09,Siko:09} and may thus be able to explain how
particle dominated plasmas are created in jets in magnetic field
dominated jets. Furthermore, the creation of mini-jets with modest bulk
Lorentz factors inside a jet with a modest bulk Lorentz factor can
explain the formation of mini-jets with very high effective bulk Lorentz
factors. This scenario would avoid problems associated with very high
bulk Lorentz factors of the jet itself, namely problems concerning the
statistics of detected objects \citep{Henr:06} and concerning tension
between high inferred Lorentz factors on the order of 50 and rather slow
observed pattern speeds in radio interferometric observations (e.g. 
\citet{Pine:10}). Future work on reconnection might corroborate this
possible link between jet formation the non-thermal emission.

\acknowledgements

VERITAS is supported by grants from the U.S. Department of Energy, the
U.S. National Science Foundation and the Smithsonian Institution, by
NSERC in Canada, by Science Foundation Ireland and by the STFC in the
U.K. We acknowledge the excellent work of the technical support staff at
the FLWO and the collaborating institutions in the construction and
operation of the instrument. The authors are grateful to the RXTE SOF
and GOF. The Metsahovi team acknowledges the support from the Academy of
Finland. The UMRAO is funded by the NSF and the University of Michigan.
Swift/BAT transient monitor results provided by the Swift/BAT team. We
also acknowledge Swift/XRT monitoring program and target of opportunity
support provided by the Swift Team and supported by NASA grants
NNX08AV77G and NNX08AT31G. We gratefully acknowledge useful discussions
with Erik Hoversten regarding Swift/UVOT analysis. H.K and A.G.
acknowledge support from the NASA grant NNX08AZ76G for the analysis of
the Suzaku data.

\end{document}